\documentclass[journal]{IEEEtran}

\renewcommand{\baselinestretch}{1}

\usepackage{tabularx,amsfonts,amssymb,amsmath}
\usepackage{algorithmic}
\usepackage[dvips]{color}
\usepackage{textcomp}
\usepackage{soul}
\usepackage{mathptmx}
\usepackage{mathrsfs}
\usepackage{subfigure}
\usepackage{calc,ifthen}
\usepackage{multirow}
\usepackage{bm}
\usepackage{float}
\usepackage{pstricks}
\usepackage{amsthm}
\usepackage{empheq}
\usepackage{setspace}

\usepackage{psfrag}
\usepackage{pstool}
\usepackage{changes}
\usepackage{mathtools}
\usepackage{bbm}
\usepackage{bm}
\usepackage{enumerate}
\usepackage{breqn}
\usepackage{array}
\usepackage{cite}
\usepackage{cuted}
\usepackage{gensymb}
\usepackage{color, colortbl}
\usepackage{ulem}

\newtheorem{rem}{\textit{Remark}}

\newcommand{\be}{\begin{equation}}
\newcommand{\ee}{\end{equation}}
\newcommand{\ben}{\begin{equation*}}
\newcommand{\een}{\end{equation*}}
\newcommand{\bea}{\begin{eqnarray}}
\newcommand{\eea}{\end{eqnarray}}
\newcommand{\bean}{\begin{eqnarray*}}
\newcommand{\eean}{\end{eqnarray*}}

\renewcommand{\H} {{\mathcal H}}

\newcommand{\B} {{\mathcal B}}

\newcommand{\N} {{\mathcal N}}

\newcommand{\R} {{\mathbb R}}
\newcommand{\D} {{\mathbb D}}

\newcommand{\K} {{\hat K}}

\renewcommand{\d} {{\delta}}

\newcommand{\mynote}[1]{}
\newcommand{\newnote}[1]{}
\newcommand{\cancelled}[1]

\title{\LARGE \bf
LMI-based robust model predictive control for a quarter car with series active variable geometry suspension
}
\author{Zilin Feng, Anastasis Georgiou, Simos A. Evangelou, Min Yu, Imad M Jaimoukha and Daniele Dini% <-this % stops a space
\thanks{Z. Feng ({\tt\small
zilin.feng17@imperial.ac.uk}), S. A. Evangelou {\tt\small (s.evangelou@imperial.ac.uk)} and I. M. Jaimoukha{\tt\small (i.jaimouka@imperial.ac.uk)} are with the Dept. of Electrical and Electronic Engineering at Imperial College London, UK.}
\thanks{ A. Georgiou ({\tt\small
georg611@umn.edu}) is with the Dept. of Mechanical Engineering at University of Minnesota, USA.}
\thanks{M. Yu {\tt\small
(m.yu14@imperial.ac.uk)} and D. Dini {\tt\small
(d.dini@imperial.ac.uk)} are with the Dept. of Mechanical Engineering at Imperial College London, UK.}}
\begin{document}
%\bstctlcite{IEEEexample:BSTcontrol}
\maketitle
%\thispagestyle{empty}
%\pagestyle{empty}   systematic tuning process 
%%%%%%%%%%%%%%%%%%%%%%%%%%%%%%%%%%%%%%%%%%%%%%%%%%%%%%%%%%%%%%%%%%%%%%%%%%%%%%%%
\begin{abstract}
This paper proposes a robust model predictive control-based solution for the recently introduced series active variable geometry suspension (SAVGS) to improve the ride comfort and road holding of a quarter car. In order to close the gap between the nonlinear multi-body SAVGS model and its linear equivalent, a new uncertain system characterization is proposed that captures unmodeled dynamics, parameter variation, and external disturbances. Based on the newly proposed linear uncertain model for the quarter car SAVGS system, a constrained optimal control problem (OCP) is presented in the form of a linear matrix inequality (LMI) optimization. More specifically, utilizing semidefinite relaxation techniques a state-feedback robust model predictive control (RMPC) scheme is presented and integrated with the nonlinear multi-body SAVGS model, where state-feedback gain and control perturbation are computed online to optimise performance, while physical and design constraints are preserved. Numerical simulation results with different ISO-defined road events demonstrate the robustness and significant performance improvement in terms of ride comfort and road holding of the proposed approach, as compared to the conventional passive suspension, as well as, to actively controlled SAVGS by a previously developed conventional $H_{\infty}$ control scheme.
\end{abstract}

%%%%%%%%%%%%%%%%%%%%%%%%%%%%%%%%%%%%%%%%%%%%%%%%%%%%%%%%%%%%%%%%%%%%%%%%%%%%%%%%
\section{INTRODUCTION}\label{sec:Intro}
The suspension system refers to the entire support system composed of springs and shock absorbers between the vehicle body and wheels. The function of the suspension system is to isolate passengers from vibration and shocks induced by road disturbances to improve ride comfort while providing good road holding \cite{jazar2017vehicle}. The passive suspension includes a conventional spring-damper unit, which passively adapts to the road profile and dissipates energy from road perturbations. Active suspension systems have become widely available and popular since the 1980s. With improved ride comfort as compared to passive suspension, concise structure, reduced energy costs and high reliability, they are continuously pursued in the process of vehicle and transportation electrification.

\begin{figure}[h]
\begin{center}
\includegraphics[width=1\columnwidth]{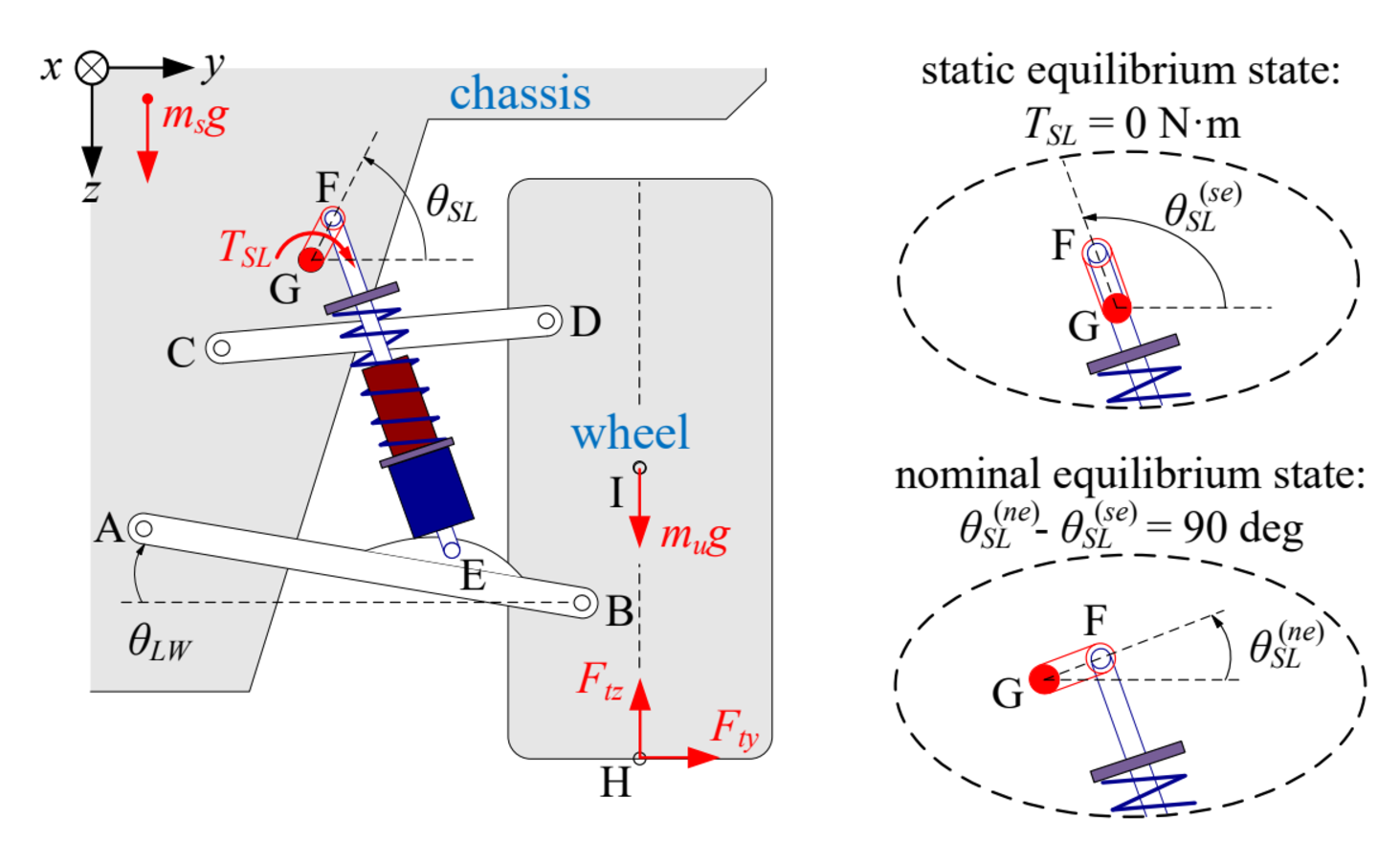}    % The printed column width is 8.4 cm.
\vspace{-2mm}
\caption{SAVGS application to a quarter car double-wishbone suspension~\cite{yu2017quarter}. $\theta_{SL}$ denotes the single-link angle with respect to the horizontal plane (as shown in the diagram it has a negative value). The superscript $(se)$ refers to the static equilibrium state with zero torque $T_{SL}=0$ applied on the single link. The superscript $(ne)$ denotes the nominal equilibrium state, where $\theta_{SL}^{(ne)}-\theta_{SL}^{(se)}=\Delta\theta_{SL}^{(ne)}=90\degree$ (Note that $\theta_{SL}-\theta_{SL}^{(se)}=\Delta\theta_{SL}$).}
\label{fig1-1}
\end{center}
\end{figure}

Recently, a novel mechatronic suspension solution, the Series Active Variable Geometry Suspension (SAVGS), has been conceptualised, designed, optimised, and experimentally validated through both quarter-car prototyping and full-car road testing~\cite{arana2016series,arana2017series,yu2017quarter,Aranaphdthesis,cheng2019series,yu2019robust,yu2021series}. Fig. \ref{fig1-1} shows the SAVGS application to a quarter-car with a double-wishbone suspension, where the active single-link component (`F-G') is introduced between the chassis (`G') and the upper-end eye of the spring damper unit ('F'), in series with the conventional spring-damper. The single-link (SL) is driven by a rotary permanent magnet synchronous motor (PMSM) actuator, which generates the torque ($T_{SL}$) acting from the chassis onto the lower wishbone (through the spring-damper) to improve the performance of a double-wishbone suspension. As compared with other active suspensions, the SAVGS features~\cite{arana2016series}: i) potential for improvement of ride comfort and road holding, ii) negligible unsprung mass and small sprung mass increment, iii) low power actuation requirements, and iv) fail-safe characteristics.

\vspace{-2mm}

Previous works have employed robust control methodologies for the ride comfort and road-holding enhancement of automobile active suspension systems and achieved reasonable success. For example, the $H_{\infty}$ control has been synthesised with a quarter-car SAVGS in \cite{yu2017quarter,arana2017series} for ride comfort enhancement and the $\mu$-synthesis approach has been proposed in \cite{feng2020uncertainties} for a full car SAVGS to improve suspension performance. A $\mu$-synthesis control solution has also been proposed for a full car with another recently introduced mechatronic suspension, the Parallel Active Link Suspension (PALS) \cite{feng2022mu}. 
%Despite these achievements, the previous control design relies on linearised models of the physical system, thus it misses the opportunity to account for system nonlinearities as uncertainties, which may result in the lack of robustness and less performing results. 
Despite these achievements, the physical constraints, such as actuator torque and actuator speed constraints, are not explicitly taken into account in the previous control design. Instead by tuning associated weights in the control design it is hoped that the actuator will operate not too close to its limits. Furthermore, the previous robust control schemes do not allow for continuously adapting the control solution, instead they design offline control gains that remain fixed with time. Therefore, both of these shortcomings could lead to constraint violation and performance conservativeness. In addition, in previously developed control techniques \cite{yu2017quarter,feng2020uncertainties} it is only possible to account for the road inputs as unbounded exogenous disturbances and model miss-match as structured uncertainties, which could lead to a lack of robustness in the presence of disturbances and model uncertainties with more precise characteristics.

A very promising control approach that has gained a lot of attention in the last two decades is Model Predictive Control (MPC). MPC refers to a class of algorithms in which the current control action is computed by solving online, at each sampling instant, an optimization problem based on an explicit model of the system\cite{Mayne00, Rawlings17, Maciejowski02}. This model-based control technique has been widely implemented in industry due to its ability to directly handle any physical and/or design constraints within its formulation~\cite{Qin03}. An implementation of a standard MPC approach on an active suspension control application has been presented in \cite{mehra1997active, gohrle2013model,Yao2021, Enders2020}. Although it partially addresses some of the problems mentioned earlier (i.e. it does take into account constraints and updates the control gain online), it is not the preferred control method for a real-life implementation since it still lacks the potential to take into account model miss-match in the form of structured feedback uncertainties and/or external disturbances into the optimization problem. An explicit MPC formulation is presented in \cite{Theunissen2019}, where the OCP is designed for an ideal linear quarter-car active suspension system (known and fixed spring and damping coefficients) with a preview of road disturbances. A quadratic programming formulation of a model predictive controller based on a differential flatness derivation of the nonlinear active suspension system of a quarter car is presented in \cite{Rodriguez2023}.

The robust formulation of MPC schemes has been the subject of extensive research with many schemes in the literature (see, for example, \cite{Scokaert98, Kerrigan04,kothare1996robust, cuzzola2002improved, Garone12, tahir2013causal, fleming2014robust, HANEMA20} and the reference therein). The adaptive tube-based model predictive control is presented in \cite{kang2020adaptive} for a vehicle active suspension system, which guarantees robustness against model uncertainties and external disturbances. Tube-based MPC methods involve initially forecasting a nominal system trajectory and ensuring that every projected closed-loop state trajectory of the uncertain system remains within a "tube" around the nominal trajectory \cite{Langson04, Mayne05, bujarbaruah2021simple, fleming2014robust, Vilaivannaporn21}. One of the major advantages of tube-based MPC is that its online computational complexity is similar to the nominal MPC formulation. However, the control performance of this method is compromised due to a fixed control gain used offline to calculate the volume of tubes based on the non-trivial concept of invariant sets. An alternative RMPC formulation using LMIs has been proposed in \cite{kothare1996robust, cuzzola2002improved, tahir2013causal, Georgiou2022}, which does not involve any offline calculation, and therefore it is less conservative, while external disturbances and model uncertainties are considered. 

In this paper, an LMI-based RMPC formulation, inspired by the algorithm proposed in \cite{Georgiou2022,Anastasisphdthesis} is employed, considering a proposed uncertain linear equivalent suspension model to control a high-fidelity SAVGS quarter car system. The LMI-based RMPC is expected to have the following advantages: i) the constraints for the actuator limits can be explicitly taken into account in the formulation of the MPC scheme, ii) the state-feedback gain and control perturbation are computed online at each sampling interval to solve the optimization problem, iii) the road inputs can be considered as exogenous disturbances with pre-defined realistic bounds against which the RMPC scheme is developed, and iv) the model uncertainty can be considered in a structured feedback manner. Therefore, the first two advantages can contribute to reduced conservativeness and improved performance in comparison to other control methods, while the last two advantages can guarantee robustness under the worst-case scenario.
In addition to the RMPC controller, in this work a PI controller is also employed in parallel to the linear equivalent (and nonlinear) quarter car model, to achieve zero steady-state tracking error on the linearization point (see Section~\ref{overall:control} %\ref{subsec:PID}
for more details).

In more detail, the main contributions of this paper are: i) the development of the uncertain system to capture the model mismatch between the linear equivalent model and nonlinear high fidelity model of the quarter car SAVGS, ii) the development of a coupling control strategy, which combines a RMPC scheme with PI controller to improve suspension performance while guaranteeing the robustness and stability of the quarter car SAVGS under model uncertainties and external road disturbances, and iii) numerical simulations with a nonlinear multi-body model of the SAVGS quarter car to assess the effectiveness and robustness of the proposed control scheme, as compared to the passive suspension and the actively controlled SAVGS by state-of-the-art $H_{\infty}$ control. 

The rest of the paper is structured as follows: Section~\ref{sec:model} illustrates the nonlinear and linearised quarter-car models and describes how the uncertainties are employed in these models. Section~\ref{sec:control} designs a robust control scheme with bounded disturbances and structured feedback uncertainties taken into consideration. Section~\ref{overall:control} combines the RMPC scheme detailed in section~\ref{sec:control} with a PI controller to develop the overall scheme to improve suspension performance. Section~\ref{sec:Simulations} performs numerical simulations to compare the developed scheme to the passive suspension and the previously developed $H_{\infty}$ control scheme for the SAVGS, with ride comfort and road holding being the primary indexes. Finally, concluding remarks are discussed in Section~\ref{sec:conclusion}.

\begin{figure}[htb!]
\centering
\begin{subfigure}[]
{
%    \centering
    \includegraphics[width=0.55\columnwidth]{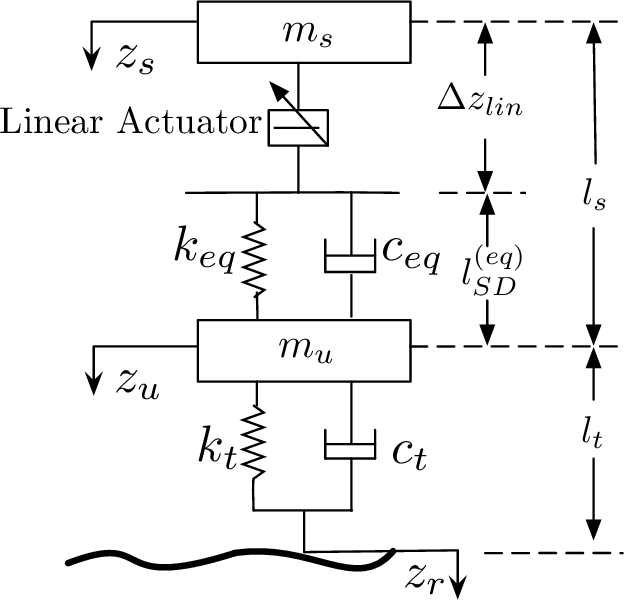}
    \label{fig2.a}}
\end{subfigure}
%\hspace{2cm}
\vspace{-3mm}
\\
%???
\begin{subfigure}[]{
%    \centering
    \includegraphics[width=0.75\columnwidth]{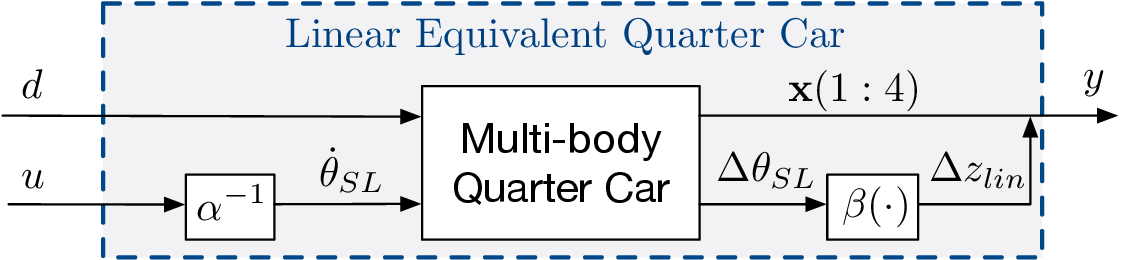}
    \label{fig2.b}}
\end{subfigure}
\vspace{-2mm}
\caption{(a) Schematic of SAVGS linear equivalent quarter car model ($z_s$ and $z_u$ are the linear equivalent displacements of the sprung and unsprung mass in the vertical direction, respectively, $z_r$ is the vertical displacement of road surface, and ${l_{{SD}}^{(eq)}}$ is the equivalent spring-damper length); (b) transformation between linear equivalent and multi-body models of the SAVGS quarter car\cite{yu2017quarter}. The function $a$ converts the control input $u=\dot{z}_{lin}$ to the rotary single-link velocity $(\omega_{SL}=\dot{\theta}_{SL})$, and the function $\beta$ converts the SL angle $\Delta \theta_{SL}$ to the linear actuator displacement $\Delta z_{lin}$. $d$ is the derivative of the road displacement profile and is considered an exogenous unbounded disturbance, and $y$ represents the measurable system's state $x(t)=[\dot{z}_{s},\dot{z}_{u},\Delta l_{s},\Delta l_{t}, \Delta z_{lin}]^T$ .}
%\label{fig1-2}
\end{figure}
\vspace{-2mm}

\subsection{Notation}\label{symbol_notation}
$\mathbb{R}$ denotes the set of real numbers, $\mathbb{R}^n$ denotes the space of $n$-dimensional real (column) vector. $\mathbb{R}^{n \times m}$ denotes the space of $n\times m$ real matrices and
$\mathbb{D}^n$ denotes the space of diagonal matrices in $\R^{n\times n}$. $\mathcal{H}(A)\!:\!=A\!+\!A^T$ for $A \!\in\! \mathbb{R}^{n \times n}$ and $A^T$ denotes the transpose of $A$. If ${\pmb{V}}\!\subseteq\!\R^{p\times q}$ is a subspace, then ${\bm\B}{\pmb{V}}\!=\!\{V\!\in\! \pmb{V}: VV^T\preceq\! 1\}$  denotes the unit ball of $\pmb{V}$.

\section{Suspension model and design requirements}\label{sec:model}
This section summarises a high-fidelity nonlinear multi-body quarter car model developed in \cite{yu2017quarter} that will be applied for nonlinear simulations and evaluation. Then the nominal linear equivalent model of the quarter car model is briefly described. The nominal linear equivalent model is initially presented in \cite{yu2017quarter,arana2017series} and proved to be sufficiently accurate for developing a nominal controller, which can then be effectively utilized for controlling the vehicle's SAVGS suspension. As mentioned in Section~\ref{sec:Intro}, the main two drawbacks of considering the nominal linear equivalent model for control design are the lack of robustness guarantees under uncertainties and the introduction of conservativeness in the control performance.
%In addition, for the implementation of the controllers, the linear equivalent model of the quarter car SAVGS derived in \cite{yu2017quarter,arana2017series} is briefly stated and proved to be accurate enough to control the vehicle suspension problem. 
To close the gap between the nonlinear and the nominal linear quarter car model, an uncertain linear equivalent model subject to parameter variation and external disturbances is derived to capture the nonlinear dynamics of the actual system.
%Finally, by characterizing the uncertainties in the linear equivalent model, the uncertain system is first derived to capture the dynamics and nonlinearity of the actual system.
\vspace{-2mm}
\subsection{Nonlinear multi-body model of quarter car SAVGS}\label{sec:2-1}
The nonlinear multi-body quarter car model of the SAVGS-retrofitted suspension has been defined in Autosim~\cite{sayers1993autosim,mousseau1992symbolic}, which extends the conventional quarter car to include a double-wishbone geometric arrangement and a nonlinear damper force characteristic, as shown in Fig. \ref{fig1-1}\cite{arana2017series,yu2017quarter}. It involves a sprung mass that is allowed to move vertically and an unsprung mass that is connected to it via a massless double wishbone kinematic linkage. A road tire compression force $F_{tz}$ proportional to the tire deflection, acts on the unsprung mass to support the overall mass of the quarter car and to introduce the road forcing. The main components of a nonlinear SAVGS model consisting of the SL and its PMSM actuator and gearbox are further integrated in series with the conventional spring-damper to complete the SAVGS retrofit in the quarter car model. 
\vspace{-2mm}
\subsection{Quarter Car SAVGS linear equivalent model (nominal)}\label{sec:2-2}
To enable the linear robust control synthesis, a linear equivalent model of the SAVGS quarter car derived in \cite{yu2017quarter,arana2017series}, as shown in Fig.~\ref{fig2.a}, 
is summarized here. The equivalent model is hand-derived utilizing energy-based linearization of the quarter-car SAVGS multi-body model described in Section \ref{sec:2-1}, by also removing the main geometric nonlinearity associated with the SL rotation, as shown in Fig.~\ref{fig2.b}.
The suspension geometric nonlinearity associated with the SL angle variation ($\Delta \theta_{SL}$) is lumped into the functions $\alpha$ and $\beta$ %($=\dot{z}_{lin}/\dot{\theta}_{SL}$) 
such that the linear equivalent model continues to be accurate for a large range of operating conditions in $\Delta \theta_{SL}$ ($\alpha$ and $\beta$ are precisely defined later in section \ref{overall:control}). 
The equations of motion of the quarter car SAVGS linear equivalent model are:
\begin{equation}\label{motion_eq}
\begin{aligned}
m_s\ddot{z}_s &= k_{eq}(\Delta{l}_s-\Delta{z}_{lin}) + c_{eq}(\dot{l}_{s}-\dot{z}_{lin}) \\
m_u\ddot{z}_{u} &= -k_{eq}(\Delta{l}_s-\Delta{z}_{lin}) - c_{eq}(\dot{l}_{s}-\dot{z}_{lin}) + k_t\Delta{l}_{t} + c_t\dot{l}_t.
\end{aligned}
\end{equation}
which can be written in state space form as follows:
\begin{equation}\label{1-1}
\begin{aligned}
\dot{x}(t) &= Ax(t) + B\bar{u}(t), \\
\end{aligned}
\end{equation}
where the system state is $x(t)=[\dot{z}_{s},\dot{z}_{u},\Delta l_{s},\Delta l_{t}, \Delta z_{lin}]^T$, in which: a) $\dot{z}_{s}$ is the sprung mass vertical velocity, b) $\dot{z}_{u}$ is the unsprung mass vertical velocity, c) $\Delta{l}_{s}$ is the increment of the suspension deflection $l_s={z}_{u} - {z}_{s}$ from its nominal equilibrium state (with SL at the $\Delta\theta_{SL}=90\degree$ position), 
d) $\Delta{l}_{t}$ is the increment of the tire deflection $l_t={z}_{r} - {z}_{u}$ from its nominal equilibrium state (with SL at the $\Delta\theta_{SL}=90\degree$ position) 
and where $z_s$, $z_u$, $z_r$ are the displacements of sprung mass, unsprung mass and road, respectively, and e) $\Delta{z}_{lin}$ %$={z}_{lin}-{z}_{lin}^{(ne)}$ 
is the displacement increment of the linear actuator with respect to its nominal equilibrium state (with SL at the $\Delta\theta_{SL}=90\degree$ position). The system input vector is $\bar{u}(t)=[d,u]^T$, where $d=\dot{z}_{r}$ is the derivative of the road displacement profile taken as an exogenous unbounded disturbance, and $u=\dot{z}_{lin}$, which is the control input, is the derivative of the linear equivalent actuator displacement increment ($\Delta{z}_{lin}$) (see Fig. \ref{fig2.a}). 

The matrices in \eqref{1-1} are
\begin{equation}\label{eq:unc}
\begin{aligned}
&A = 
\begin{bmatrix}
-\frac{c_{eq}}{m_{s}}\,& \frac{c_{eq}}{m_{s}}\,& \frac{k_{eq}}{m_{s}}\,& 0\,& -\frac{k_{eq}}{m_{s}}\\ 
\frac{c_{eq}}{m_{u}}\,& -\frac{c_{eq}+c_t}{m_{u}}\,& -\frac{k_{eq}}{m_{u}}\,& \frac{k_{t}}{m_{u}}\,& \frac{k_{eq}}{m_{u}}\\ 
-1\,& 1\,& 0\,& 0\,& 0 \\
0\,& -1\,& 0\,& 0\,& 0 \\
0\,& 0\,& 0\,& 0\,& 0 \\
\end{bmatrix},\,~B = \begin{bmatrix}
-\frac{c_{eq}}{m_{s}}\,& 0\\
\frac{c_{eq}}{m_{u}}\,& \frac{c_{t}}{m_{u}}\\
0\,& 0\\
0\,& 1\\
1\,& 0\\
\end{bmatrix}
\end{aligned}
\end{equation}
%& B^T = 
%\begin{bmatrix}
%-\frac{c_{eq}}{m_{s}} & \frac{c_{eq}}{m_{u}} & 0 & 0 & 1 \\
%0 & \frac{c_{t}}{m_{u}} & 0 & 1 & 0
%\end{bmatrix},
where $m_s$ and $m_{u}$ are the sprung and unsprung masses respectively, $k_{eq}$ and $c_{eq}$ are the equivalent suspension stiffness and damping coefficients (from the passive spring-damper), respectively, and $k_{t}$ and $c_{t}$ are the tire stiffness and damping coefficients, respectively. 

\subsection{Quarter Car SAVGS linear equivalent uncertain model}\label{sec:2-3}
Nonlinearities and unmodeled dynamics of the actual suspension system can be approximately captured by utilizing an uncertain linearized model,  which provides a more reliable basis for designing a robust controller~\cite{feng2020uncertainties}. Therefore, this work aims to close the gap between the nonlinear system and its equivalent linear version presented in Section~\ref{sec:2-2}. The first step for doing so is to consider the signal $d$ as a bounded road disturbance signal with defined bounds that will be considered in the control design to preserve robustness, instead of an exogenous unbounded signal. It is assumed that the road profile has symmetric upper and lower bounds, with $d\in [-\bar{d},\bar{d}]$ where $\bar{d}$ is a positive number representing the largest road profile amplitude (see more details about $\bar{d}$ later in subsection~\ref{sec:Simulations}). Then, it is reasonable to consider the suspension damping coefficient $c_{eq}$ as a (time-invariant) uncertain parameter to capture its nonlinear characteristic within its operating speed range; see damper manufacturer datasheets  and \cite{feng2022mu} for damper nonlinear characteristic. %the relationship between the operating speed range and the damping ratio, which have nonlinear speed-dependent characteristics. 
Its variation from the nominal (low-damper-speed) value can be captured by introducing the time-invariant norm-bounded structured uncertainty operator $\Delta$ and the uncertain input and output signals $p(t)$ and $q(t)$, respectively. Therefore, the new SAVGS quarter car model %suspension system 
subject to additive disturbances and structured feedback uncertainties can be described as follows:
\begin{equation}\label{1-2}
\begin{aligned}
\dot{x}(t) &= Ax(t) + B_{u}u(t) + B_{p}p(t) + B_{d}d(t),\\
q(t)&=C_q x(t)+D_{qu} u(t),\\
p(t)&=\Delta q(t).
\end{aligned}
\end{equation}
The matrices in \eqref{1-2} are:
\begin{equation} \label{matrix_unc}
\begin{aligned}
&A = 
\begin{bmatrix}
-\frac{c_{eq}^{(nom)}}{m_{s}} & \frac{c_{eq}^{(nom)}}{m_{s}} & \frac{k_{eq}}{m_{s}} & 0 & -\frac{k_{eq}}{m_{s}}\\
\frac{c_{eq}^{(nom)}}{m_{u}} & -\frac{c_{eq}^{(nom)}+c_t}{m_{u}} & -\frac{k_{eq}}{m_{u}} & \frac{k_{t}}{m_{u}} & \frac{k_{eq}}{m_{u}}\\ 
-1 & 1 & 0 & 0 & 0 \\
0 & -1 & 0 & 0 & 0 \\
0 & 0 & 0 & 0 & 0 \\
\end{bmatrix},\\
& B_{u}^T = 
\begin{bmatrix}
-\frac{c_{eq}^{(nom)}}{m_{s}} & \frac{c_{eq}^{(nom)}}{m_{u}} & 0 & 0 & 1 
\end{bmatrix},\\
& B_{d}^T = 
\begin{bmatrix}
0 & \frac{c_{t}}{m_{u}} & 0 & 1 & 0
\end{bmatrix},~~~ B_{p}^T = 
\begin{bmatrix}
\frac{1}{m_{s}} & -\frac{1}{m_{u}} & 0 & 0 & 0
\end{bmatrix},\\
& C_q =\begin{bmatrix}
-c_{eq}^{(dev)} & c_{eq}^{(dev)} & 0 & 0 & 0
\end{bmatrix},~~~D_{qu}=\begin{bmatrix}
-c_{eq}^{(dev)}
\end{bmatrix},\\
\end{aligned}
\end{equation}
in which $c_{eq}^{(nom)}$ and $c_{eq}^{(dev)}$ ($c_{eq}^{(dev)}>0$) denote the nominal value and the maximum deviation of suspension damping coefficient $c_{eq}$, respectively, where:
\begin{equation}
\begin{aligned}
c_{eq} = c_{eq}^{(nom)} \pm  c_{eq}^{(dev)}.
\end{aligned}
\end{equation}
Taking into account the operational damper speed range, it is assumed that $c_{eq}$ %, presented on the linear equivalent model, 
can be within a $\pm10\%$ range of its nominal value, therefore,  %($c_{eq}^{(min)}\leq c_{eq}\leq c_{eq}^{(max)}$), where $c_{eq}^{(max)}=1.1c_{eq}^{(nom)}$ and $c_{eq}^{(min)}=-1.1c_{eq}^{(nom)}$. Based on these bounds, the maximum deviation of the damping coefficient is defined as 
$c_{eq}^{(dev)}=0.1c_{eq}^{(nom)}$.
%0.5 (c_{eq}^{(max)}-c_{eq}^{(min)})$. 
The uncertainty operator has the form $\Delta\!:=\!\{\Delta\in \mathbb{R}: \Delta^T\Delta\leq 1\}$ and the disturbance set is bounded $\mathcal{D} \!:=\! \{ d \!\in\! \mathbb{R}: -\bar{d}\!\leq\!
d\leq\! \bar{d}\}$.

Since the robust control design presented subsequently in Section~\ref{sec:control} is developed using a discrete-time model, the continuous-time model in \eqref{1-2} is discretized using a zero-order hold method, where the discrete model is defined as:
\begin{equation}\label{1-3}
\begin{aligned}
x_{k+1} &= Ax_k + B_{u}u_k + B_{p}p_k + B_{d}d_k,\\
q_k&=C_q x_k+D_{qu} u_k,\\
p_k&=\Delta_k q_k.
\end{aligned}
\end{equation}
\begin{rem}Note that the distribution matrices $A,B_u,B_p$, and so on, in \eqref{1-3} are the discretized versions of those in \eqref{1-2} and the same notation is used for simplicity.
\end{rem}
\begin{rem}In general the norm-bounded structured uncertainty operator $\Delta$ can be a time-varying parameter and can be represented as $\Delta_k$. In the present work, it is assumed that the operator is time-invariant, thus $\Delta_k=\Delta$ for all $k$. 
\end{rem}

\subsection{Objectives and constraints requirements}\label{sec:2-4}
In this study, ride comfort and road holding are selected as the two main control objectives in suspension design~\cite{sharp1986evaluation,vongierke1975iso,sharp1987road}. %add references.
As it is common in the context of the quarter-car, these two aspects of suspension performance are addressed by minimizing the vertical body acceleration of the sprung mass, $\ddot{z}_{s}$, and the vertical tire deflection, $\Delta{l}_{t}$~\cite{Aranaphdthesis}. %Hence, the vertical body acceleration $\ddot{z}_{s}$ should be minimized to isolate passengers from vibrations and improve ride comfort. To achieve the desired road-holding property, it is recommended to guarantee firm uninterrupted contact of the wheels with the road and keep the tire deflection $\Delta{l}_{t}$ as small as possible. In contrast, 
Due to structural limitations and the physical capabilities of the actuator, three hard constraints are also used in this work for the single-link angle $\Delta \theta_{SL}$, the single-link angular velocity  $\omega_{SL}=\dot{\theta}_{SL}$ (which is the control input in the nonlinear multibody quarter-car model), and the actuator single link torque $T_{SL}$ (see Section~\ref{subsec:ConstrainsAssociation} for more details on the constraint association), respectively. %In other words, these three variables should not exceed their predefined limits. 
Therefore, to achieve acceptable performance, the weighted quadratic values of the vertical body acceleration and the tire deflection are minimized, with the values of $\theta_{SL}$, $\omega_{SL}$ and $T_{SL}$ satisfying their expected bounds.

\section{Robust Model Predictive Control Design}\label{sec:control}
This section summarizes the RMPC theoretical technique in \cite{Georgiou2022}, which is utilized in the proposed causal state-feedback RMPC methodology for the SAVGS quarter car control problem. %is presented (see \cite{Georgiou2022} for %further theoretical details of the general method). 
The general form of a system description which includes control dynamics, constraints, and cost signal is first provided. Then the algebraic formulation of an online and an offline controller, which are applied to steer a system to an admissible reference signal is explained. Considering a causal state feedback control law, the optimization problem aims to compute a state-feedback gain and a control perturbation by solving linear matrix inequalities, where the computational burden is substantially reduced without adversely affecting the tracking performance. An offline strategy to guarantee feasibility of the RMPC problem is also employed. Finally, the overall RMPC algorithm utilized in this paper is presented to summarize the control strategy that is followed in the proposed methodology.

\subsection{System Description}
Following the representation described in Section~\ref{sec:2-3}, the general form of the linear discrete-time system, subject to bounded disturbances and norm-bounded structured uncertainty is illustrated as~\cite{kothare1996robust}:
\vspace{-1mm}
\begin{equation}\label{eq:dt_system_model}
\begin{aligned}
\hspace{-1mm}\left.\begin{array}{c}~\\
\begin{bmatrix}
x_{k+1} \\ q_k \\ f_k \\ z_k
\end{bmatrix}\end{array}\right.\!\!\!\!\!\!\!\!\!&=\!\!\!\!\!\!\!\!\!\!\!\left.\begin{array}{rl}~&\!\!\left.\!\!\!\!\!\!\!\!\!\begin{array}{cccc}\scriptstyle{n}&~~\scriptstyle{n_u}&~~\scriptstyle{n_p}&~~~\scriptstyle{n_d}\end{array}\right.\\
\left.\begin{array}{c}\scriptstyle{n}\\\scriptstyle{n_q}\\\scriptstyle{n_f}\\\scriptstyle{n_z}\end{array}\right.&\!\!\!\!\!\!\!\!\!\!\!\!\!\!\begin{bmatrix}
A     & B_u & B_p & B_d\\
C_q & D_{qu} &0&0 \\
C_f & D_{fu}  & D_{fp} & D_{fd}\\
C_z & D_{zu} & D_{zp} & D_{zd}
\end{bmatrix}\end{array}\right.\!\!\!\!\!\!\!\!\!\!\!\!\!\!\left.\begin{array}{c}~\\
\begin{bmatrix}
x_k\\u_k\\p_k\\d_k
\end{bmatrix}\end{array}\right.\!\!\!\!\!\!\!\!\!,\!\!\!\!\!\!&p_k&\!=\!\Delta_k q_k, \\
\begin{bmatrix}
 q_N \\ f_N \\ z_N
\end{bmatrix}\!\!&=\!\!\begin{bmatrix}
\hat{C}_q & 0\\
\hat{C}_f &  \hat{D}_{fp}\\
\hat{C}_z & \hat{D}_{zp}
\end{bmatrix}\!\!\begin{bmatrix}
x_N\\p_N
\end{bmatrix}\!\!, &p_N&=\Delta_N q_N, \\
\end{aligned}
\end{equation}
where $x_k \in \mathbb{R}^n$, $u_k \in \mathbb{R}^{n_u}$, $d_k \in \mathbb{R}^{n_d}$, $f_k \in \mathbb{R}^{n_f}$, $z_k \in \mathbb{R}^{n_z}$, $p_k \in \mathbb{R}^{n_p}$ and $q_k \in \mathbb{R}^{n_q}$ are the state, input, disturbance, constraint, cost, and input and output uncertainty vectors, respectively, with $k\!\in\!\N\!:=\!\{0,1,\ldots,N\!-\!1\}$, where $N$ is the horizon length. In this study, it is assumed that all the states are measurable and the description includes terminal cost signal and state constraints to ensure closed-loop stability~\cite{Rawlings17}. $\Delta_k\!\in\! \bm\B \pmb{\Delta}$ where $\pmb{\Delta}\!\subseteq\!\R^{{n_p}\times{n_q}}$ is a subspace that captures the uncertainty structure. Finally, the disturbance $d_k$ is assumed to belong to the set $\mathcal{D}_k \!=\! \{ d_k \!\in\! \mathbb{R}^{n_d} \!\!: -\bar{d}_k\!\leq\! d_k \!\leq\! \bar{d}_k\}$, where the disturbance's upper bound is $\bar{d}_k\!>\!0$ and assumed known or approximated by the application specification (see Section~\ref{sec:Simulations} for example).
\begin{rem}
The constrain signal $f_k$ and the cost signal $z_k$ are design signals linearly associated with the state and the input signal of the linear equivalent model and are used at the RMPC formulation, %(more details and design decision can be found in Sections~\ref{subsec:2-B} and \ref{sec:Simulations}).    
as will be presented in Sections~\ref{subsec:2-B}, \ref{overall:control} and \ref{sec:Simulations}.
\end{rem}
\subsection{Problem formulation} \label{subsec:2-B}
Given the initial state $x_0$, the design of the robust model predictive controller for all $k\in\N$ leads to the problem of finding a feedback law $u_k$ over the horizon $N$ such that the cost function 
\begin{equation}\label{eq:costfunction}
\begin{aligned}
J = \max_{d\in \mathcal{D}_k,\hat{\Delta}\in \mathcal{B}\pmb{\Delta}} \hspace{0.1cm} \sum_{k=0}^N (z_k-\bar{z}_k)^T(z_k-\bar{z}_k),
\end{aligned}
\end{equation} 
is minimised, while the future predicted outputs satisfy the constraints $f_k\leq\bar{f}_k$ and $f_N\leq\bar{f}_N$ for all $d_k \in \mathcal{D}_k$ and all $\Delta_k\in\mathcal{B}\pmb{\Delta}$ and for all $k\in N$.
The parameter $\bar{z}_k$ defines the reference trajectory and $\bar{f}_k$ and $\bar{f}_N$ are chosen to include polytopic constraints on input, state and output  signals, and terminal signals respectively.

To simplify the presentation, the disturbance is re-parameterised as uncertainty by redefining $\mathcal{D}_k\!:=\!\{\Delta_k^d\bar{d}_k\!\!:\Delta_k^d\!\in\!\bm\B\pmb{\Delta}^d\}$, where $\pmb{\Delta}^d\!=\!\mathbb{D}^{n_d}$ and, 
\begin{equation*}
B_p\!:=\!\!\left[\!\!\!\!\begin{array}{cc}B_p&\!\!\!\!B_d\end{array}\!\!\!\!\right]\!\!\!,C_q\!:=\!\!\left[\!\!\!\!\begin{array}{c}C_q\\0\end{array}\!\!\!\!\right]\!\!\!,D_{qu}\!:=\!\!\left[\!\!\!\!\begin{array}{c}D_{qu}\\0\end{array}\!\!\!\!\right]\!\!\!,
\bar{d}_k\!:=\!\!\left[\!\!\!\!\begin{array}{c}0\\\bar{d}_k\end{array}\!\!\!\!\right]\!\!\!,p_k\!:=\!\!\left[\!\!\!\!\begin{array}{c}p_k\\d_k\end{array}\!\!\!\!\right]\!\!\!,
\end{equation*}
$q_k\!:=\!C_qx_k+D_{qu}u_k+\bar{d}_k$
and $n_p\!:=\!n_p\!+\!n_d,n_q\!:=\!n_q\!+\!n_d.$ 

By defining the stacked vectors,
\begin{equation*}
    \mathbf{u}=\!\!\left[\!\!\begin{array}{c}u_0\\\vdots\\u_{N-1}\end{array}\!\!\right]\!\!\in \mathbb{R}^{N_u},~\mathbf{x}=\!\!\left[\!\!\begin{array}{c}x_1\\\vdots\\x_N\end{array}\!\!\right]\!\!\in \mathbb{R}^{N_n},~\pmb{\zeta}=\!\!\left[\!\!\begin{array}{c}\zeta_0\\\vdots\\\zeta_N\end{array}\!\!\right]\!\!\in \mathbb{R}^{N_\zeta},
\end{equation*}
where $\pmb{\zeta}$ stands for $\mathbf{f},\bar{\mathbf{f}},\mathbf{p},\mathbf{q},\mathbf{z},\bar{\mathbf{z}}$ or $\bar{\mathbf{d}}$ and $N_n=\!Nn,~\!N_u\!=\!Nn_u$ and $N_\zeta\!=\!(N\!+\!1)n_\zeta$, we get
\begin{equation}
\label{Stacked}
\left.\begin{array}{c}~\\
\begin{bmatrix}
\mathbf{x}\\ \mathbf{q} \\ \mathbf{f} \\ \mathbf{z}
\end{bmatrix}\end{array}\right.\!\!\!\!\!\!\!\!\!=\!\!\!\!\!\!\!\!\!\!\!\left.\begin{array}{rl}~&\!\!\left.\!\!\!\!\!\!\!\!\!\begin{array}{cccc}\scriptstyle{n}&~~\!\scriptstyle{N_u}&~~\scriptstyle{N_p}&~~\!\scriptstyle{1}\end{array}\right.\\
\left.\begin{array}{c}\scriptstyle{N_n}\\\scriptstyle{N_q}\\\scriptstyle{N_f}\\\scriptstyle{N_z}\end{array}\right.&\!\!\!\!\!\!\!\!\!\!\!\!\!\!
\begin{bmatrix}
\mathbf{A} & \mathbf{B}_u & \mathbf{B}_p & 0 \\
\mathbf{C}_q  &\mathbf{D} _{qu} & \mathbf{D} _{qp} & \bar{\mathbf{d}} \\
\mathbf{C}_f  &\mathbf{D} _{fu} & \mathbf{D} _{fp} & 0 \\
\mathbf{C}_z  &\mathbf{D} _{zu} & \mathbf{D} _{zp} & 0 \\
\end{bmatrix}\end{array}\right.\!\!\!\!\!\!\!\!\!\!\!\!\!\!\left.\begin{array}{c}~\\
\begin{bmatrix}
x_0 \\ \mathbf{u} \\ \mathbf{p} \\ 1
\end{bmatrix}\end{array}\right.\!\!\!\!\!\!,\quad \mathbf{p} = \hat\Delta \mathbf{q},
\end{equation}

with $\hat\Delta \in \bm\B \pmb{\hat{\Delta}}\subset\R^{N_p\times N_q} $ where,
\bean
    \pmb{\hat\Delta}\!=\!\{{\rm diag}(\Delta_0,\Delta_0^d,\ldots,\Delta_{N\!-\!1},\Delta_{N\!-\!1}^d,\Delta_N)\!\!:\!\Delta_k\!\in\!\pmb{\Delta},\Delta_k^d\!\in\!\pmb{\Delta}^d \}\!,
\eean
and where the stacked matrices in \eqref{Stacked} (shown in bold) have the indicated dimensions and are readily obtained from iterating the dynamics in \eqref{eq:dt_system_model}.

The input signal $u_i$ is considered as a causal state feedback that depends only on states $x_0,\ldots,x_i$ (see e.g. \cite{Skaf}). Thus
\begin{equation}\label{u1}
\mathbf{u}=K_0 x_0+K\mathbf{x}+\pmb{\upsilon},
\end{equation}
where $\pmb{\upsilon} \!\in \!\mathbb{R}^{N_u}$ is the (stacked) control perturbation vector and
$K_0,~\!K$ are the current and predicted future state feedback gains. Causality is preserved by restricting $[ K_0 ~\! K] \!\in\!\mathcal{K}\!\subset\! \mathbb{R}^{N_u \times N_n}$, where $\mathcal{K}$ is lower block diagonal with $n_u\times n$ blocks. 

Using the definition of $\textbf{x}$ in equation \eqref{Stacked} the control law in \eqref{u1} can be rewritten as:
\begin{equation}\label{eq:u2}
\mathbf{u}\!=\!\hat{K}_0x_0\!+\!\hat{K}\mathbf{B}_p\mathbf{p}\!+\!\hat{\upsilon},
\end{equation} 
where 
\begin{equation}
\begin{bmatrix}\hat{K}_0 & \!\!\hat{K} & \!\!\hat{\upsilon}
\end{bmatrix}\!\!=\! (I\!-\!K\mathbf{B}_u)^{-1}
\!\begin{bmatrix}
K_0\!+K\mathbf{A} & \!\!K & \!\!\pmb{\upsilon}
\end{bmatrix}.    
\end{equation}
Note that $(I\!-\!K\mathbf{B}_u)$ is invertible due to the lower-triangular structure and that $\mathbf{u}$ is affine in $\hat{K}_0,\hat{K}$ and $\hat{\pmb{\upsilon}}$ which have the same structure as ${K_0,K}$ and $\pmb{\upsilon}$. A standard feedback re-parameterization gives%\vspace{-4mm}
\begin{equation}\label{eq:Kv}
\begin{bmatrix}
 K_0 & K & \pmb{\upsilon}
\end{bmatrix}= (I+\hat{K}\mathbf{B}_u)^{-1}
\begin{bmatrix}
\hat{K}_0\!-\!\hat{K}\mathbf{A} & \hat{K} & \hat{\pmb{\upsilon}}
\end{bmatrix}\!\!,
\end{equation} 
and therefore $\left[\hat{K}_0~\hat{K}~\hat{\upsilon}\right]$ will be used as the decision variables instead.
Using (\ref{eq:u2}) to eliminate $\mathbf{u}$ from (\ref{Stacked}) and re-arranging $x_0$ gives
\begin{equation}\label{def1}
\begin{split}
\hspace{-3mm}\left[\!\!\!\begin{array}{c}\mathbf{q}\\\mathbf{f}\\\mathbf{z}-\bar{\mathbf{z}}\end{array}\!\!\!\right]\!\!\!&=\!\!\!\left[\!\!\!\begin{array}{c|c}
\mathbf{D}_{qp}^{\hat{K}}\!\!\!&\!\!\!\mathbf{D}_{q}^{\hat{K}_0,\hat{\upsilon}}\\ \hline
\mathbf{D}_{fp}^{\hat{K}}\!\!\!&\!\!\!\mathbf{D}_{f}^{\hat{K}_0,\hat{\upsilon}}\\ \hline
\mathbf{D}_{zp}^{\hat{K}}\!\!\!&\!\!\!\mathbf{D}_{z}^{\hat{K}_0,\hat{\upsilon}}\end{array}\!\!\!\right]\!\!\!
\left[\!\!\!\begin{array}{c}\mathbf{p}\\1\end{array}\!\!\right]\!\!, \\
&:=\!\!\!
\left[\!\!\!\begin{array}{c|c}
\mathbf{D}_{qp} \!\!+\!\! \mathbf{D}_{qu}\hat{K}\mathbf{B}_{p} \!\!\!&\!\!\! \mathbf{D}_{qu}\hat{\upsilon}\!\!+\!\! (\mathbf{C}_{q}\!\!+\!\!\mathbf{D}_{qu}\hat{K}_0)x_0\!\!+\!\! \bar{d}\\ \hline
\mathbf{D}_{fp}\!\!+\!\!\mathbf{D}_{fu}\hat{K}\mathbf{B}_{p}\!\!\! &\!\!\! \mathbf{D}_{fu}\hat{\upsilon} \!\!+\!\! (\mathbf{C}_{f}\!\!+\!\!\mathbf{D}_{fu}\hat{K}_0)x_0\\ \hline
\mathbf{D}_{zp}\!\!+\!\!\mathbf{D}_{zu}\hat{K}\mathbf{B}_{p} \!\!\!&\!\!\! \mathbf{D}_{zu}\hat{\upsilon}\!\!+\!\! (\mathbf{C}_{z}\!\!+\!\!\mathbf{D}_{zu}\hat{K}_0)x_0\!\!-\!\!\bar{z}\end{array}\!\!\!\right]\!\!\!\!
\left[\!\!\!\!\begin{array}{c}\mathbf{p}\\1\end{array}\!\!\!\right]\!\!\cdot
\end{split}
\end{equation}
Note that all the coefficient matrices in \eqref{def1} are affine in $\hat{K}_0,~\hat{K}$ and $\hat{\upsilon}$. Finally, eliminating $\mathbf{p}$ using $\mathbf{p}=\hat\Delta \mathbf{q}$ we get
\begin{equation}\label{def2}
    \left[\!\!\begin{array}{c}\mathbf{f}\\\mathbf{z}-\bar{\mathbf{z}}\end{array}\right]\!\!=\!\!\left[\!\!\begin{array}{c}
    \mathbf{D}_{f}^{\hat{K}_0,\hat{\upsilon}}+\mathbf{D}_{fp}^{\hat{K}}\hat\Delta(I-\mathbf{D}_{qp}^{\hat{K}}\hat\Delta)^{-1} \mathbf{D}_{q}^{\hat{K}_0,\hat{\upsilon}}\\
        \mathbf{D}_{z}^{\hat{K}_0,\hat{\upsilon}}+\mathbf{D}_{zp}^{\hat{K}}\hat\Delta(I-\mathbf{D}_{qp}^{\hat{K}}\hat\Delta)^{-1} \mathbf{D}_{q}^{\hat{K}_0,\hat{\upsilon}}
    \end{array}\!\!\right]\!\!\cdot
\end{equation}
For convenience, constraint and cost signals can be written as $\mathbf{f}\!=\!\mathcal{F}(\hat{K}_0,\hat{K},\hat{\upsilon},\hat\Delta)$ and $(\mathbf{z}\!-\!\bar{\mathbf{z}})^T\!(\mathbf{z}\!-\!\bar{\mathbf{z}})\!=\!\mathcal{Z}(\hat{K}_0,\hat{K},\hat{\upsilon},\hat\Delta)$ to emphasise dependence on the variables.
By following the procedure presented by~\cite{Tahir1}, the RPMC problem can be transformed to a min-max problem~\cite{Scokaert98}, where the objective is to find a feasible triple $(\hat{K}_0,\hat{K}, \hat{\upsilon})$ that solves 
\begin{equation}\label{eq:minmaxcostfun}
\mathbf{J}= \min_{(\hat{K}_0, \hat{K}, \hat{\upsilon}) \in \mathcal{U}} \hspace{0.2cm} \max_{\hat\Delta\in \bm\B\pmb{\Delta}} \mathcal{Z}(\hat{K}_0,\hat{K},\hat{\upsilon},\hat\Delta),
\end{equation}
The set $\mathcal{U}$ is defined as shown in \cite{Georgiou2022} to be the set of all feasible control variables $(\hat{K}_0,\hat{K},\hat{\upsilon})$ such that all the problem constraints are satisfied:
\begin{equation}\label{const_fbar}
\mathcal{U}\!:=\! \{ ([\hat{K}_0~\hat{K}],\hat{\upsilon})\!\in\!\mathcal{K}\!\times\!\mathbb{R}^{N_u}\!\!:\!\mathcal{F}(\hat{K}_0,\hat{K},\hat{\upsilon},\hat\Delta)\!\leq \!\bar{\mathbf{f}},
       \forall \hat\Delta \!\in\! \bm\B\pmb{\hat\Delta}\}.
\end{equation}
Since the optimization in (\ref{eq:minmaxcostfun})
is nonconvex, the semidefinite relaxation procedure presented in\cite[Lemma 1]{Tahir1}, is used by introducing an upper bound on the cost function \eqref{eq:minmaxcostfun}, defined by ${\gamma^2}$. After some matrix manipulations the inequalities $\mathcal{Z}(\hat{K}_0,\hat{K},\hat{\upsilon},\hat\Delta)\leq {\gamma^2} $ and $\mathcal{F}(\hat{K}_0,\hat{K},\hat{\upsilon},\hat\Delta)\le\bar{\mathbf{f}}$ holds for all $\hat\Delta \!\in\! \bm\B\pmb{\hat\Delta}$ if there exists a solution to the nonlinear matrix inequalities~\cite{Georgiou2022}:
\begin{eqnarray}\label{cost_LMI}
  T_1+\H(T_2\hat{K}\mathbf{B}_p T_3)&\succ& 0,\\
  \label{constr_LMI}
    T^i_1+\H(T^i_2\hat{K}\mathbf{B}_p T^i_3)&\succ& 0,~i=1,\ldots,N_f,
\end{eqnarray}
where \begin{equation*}
\begin{bmatrix}T_1&\!\!T_2\\T_3&\!\!0\end{bmatrix}\!=\!\!\!\!\!\!\!\!\!\!
   \left.\begin{array}{rl}~&\!\!\left.\!\!\!\!\!\!\!\!\!\begin{array}{ccccc}~\!\scriptstyle{N_z}&~~\scriptstyle{1}&~~~~~~~~~~~\scriptstyle{N_q}&~~~~~~~~~\scriptstyle{N_p}&~~\!\scriptstyle{N_u}\end{array}\right.\\
\left.\begin{array}{c}\scriptstyle{N_z}\\\scriptstyle{1}\\\scriptstyle{N_q}\\\scriptstyle{N_p}\\\scriptstyle{N_p}\end{array}\right.&\!\!\!\!\!\!\!\!\!\!\!\!\!\!
\left[\begin{array}{cccc|c}
    I & \!\!\mathbf{D}_{z}^{\hat{K}_0, \hat{\upsilon}} &\!\! \mathbf{D}_{zp} G^T &\!\! \mathbf{D}_{zp} S&\!\!\mathbf{D}_{zu}\\
    \ast &\!\! {\gamma^2} &\!\! (\mathbf{D}_{q}^{\hat{K}_0,\hat{\upsilon}})^T &\!\! 0&\!\!0 \\
    \ast &\!\! \ast &\!\! R\!+\!\H\!\left(\mathbf{D}_{qp}G^T\right) &\!\! \mathbf{D}_{qp}S&\!\!\mathbf{D}_{qu}\\
    \ast &\!\! \ast &\!\! \ast &\!\! S&\!\!0\\\hline
    0&\!\!0&\!\!G^T&\!\!S&\!\!0
    \end{array}\right]\end{array}\right.\!\!\!\!\!\!\!\!,
    \end{equation*}
$\begin{bmatrix}T^i_1&\!\!\!\!\!T^i_2\\
   T^i_3&\!\!\!\!\!0\end{bmatrix}\!\!=\!\! \!$
   \begin{equation*}
   \!\!\!\!\!\!\!\!\!\!
   \left.\begin{array}{rl}~&\!\!\left.\!\!\!\!\!\!\!\!\!\begin{array}{ccccc}~~~~\scriptstyle{1}&~~~~~~~~~~~~~~~~~~~~\scriptstyle{N_q}&~~~~~~~~~~~~~\scriptstyle{N_p}&~~~~~~~\scriptstyle{N_u}\end{array}\right.\\
\left.\begin{array}{c}\scriptstyle{1}\\\scriptstyle{N_q}\\\scriptstyle{N_p}\\\scriptstyle{N_p}\end{array}\right.&\!\!\!\!\!\!\!\!\!\!\!\!\!\!
\left[\!\!\!\!\begin{array}{ccc|c}
    e_i^T(\bar{\mathbf{f}}\!-\!\mathbf{D}_{f}^{\hat{K}_0,\hat{\upsilon}}) &\!\!\!\!\! (\mathbf{D}_{q}^{\hat{K}_0,\hat{\upsilon}})^T\!\!\!-\!\frac{e_i^T}{2}\mathbf{D}_{fp} G_i^T&\!\!\!\! -\frac{e_i^T}{2}\mathbf{D}_{fp} S_i&\!\!\!\! -\frac{e_i^T}{2}\mathbf{D}_{fu}\\
    \ast &\!\!\!\!  R_i\!+\!\H\!\left(\mathbf{D}_{qp}G_i^T\right)&\!\!\!\! \mathbf{D}_{qp} S_i&\!\!\!\!\mathbf{D}_{qu}\\
    \ast &\!\!\!\! \ast &\!\!\!\! S_i&\!\!\!\!0\\\hline
    0&\!\!\!\!G_i^T&\!\!\!\!S_i&\!\!\!\!0\end{array}\!\!\!\!\right]\!\!\!\end{array}\right.\!\!\!\!\!\!\!,
\end{equation*}
where $([ \hat{K}_0 ~ \hat{K}],\hat{\upsilon})\!\in\!\mathcal{K}\times\mathbb{R}^{N_u}$ and  $(S,R,G),\:(S_i,R_i,G_i)\!\in\!\hat{\Psi}$ are slack variables, where $\hat{\Psi}$ is defined as:
\begin{equation}\label{Slack_variable}
\widehat{\Psi}\!=\! \{(S,R,G)\!:S,R\!\succ 0,~\!S\Delta\!=\!\Delta R,~\!\H(\Delta G)\!=\!0~\!\forall\Delta\!\in\!\pmb{\widehat{\Delta}}\}.
\end{equation}
It follows that the relaxed RMPC problem can be summarised as:
\begin{equation} \label{finalprob}
\begin{aligned}
\hspace{-2mm}\min\{  {\gamma^2} \!:&([ \hat{K}_0 ~~ \hat{K}],\hat{\upsilon})\!\in\!\mathcal{K}\!\times\mathbb{R}^{N_u},(\ref{cost_LMI}),(\ref{constr_LMI}){\rm~are~satisfied},\\
&(S,R,G),\:(S_i,R_i,G_i)\!\in\! \widehat{\Psi},\: i \!\in\! \mathcal{N}_f\}.
\end{aligned}
\end{equation}
In this study, instead of solving multiple nonlinear matrix inequalities for the constraints as presented in \eqref{constr_LMI} (one for each of the $N_f$ constraints), a single nonlinear inequality is defined for all constraints, similarly to~\cite{Georgiou2022}. By doing so a reduction in the computational complexity and algorithm scalability can be achieved. Therefore without loss of generality, the multiple nonlinear inequalities presented in \eqref{constr_LMI} can be written as a single nonlinear inequality as shown below:
%In this study, instead of solving multiple linear matrix inequalities (LMI) for the constraints (one for each of the $N_f$ constraints), the single LMI approach for handling constraints proposed in~\cite{Georgiou2022} is utilised to achieve reduced computational complexity and improved algorithm scalability. Therefore, similar to \eqref{constr_LMI} $\mathcal{Z}(\K_0,\K,\hat{\upsilon},\hat{\Delta})\le{\gamma}^2$ and $\mathcal{F}(\hat{K}_0,\hat{K},\hat{\upsilon},\hat\Delta)\le\bar{\mathbf{f}}$ for all $\hat{\Delta}\!\in\!\bm\B\pmb{\hat{\Delta}}$  if there exist
%solutions $([\K_0~~\K],\hat{\upsilon})\in\mathcal{K}\!\times\mathbb{R}^{N_u}$, $(S,R,G), ( \tilde{S}, \tilde{R}, \tilde{G}) \in \widehat{ \Psi}$, $\mu\in\R$ and $M\in\D^{N_f}$ to \eqref{cost_LMI} and,
\begin{equation}\label{stateupperi0}   \tilde{T}_1+\H(\tilde{T}_2\hat{K}\mathbf{B}_p \tilde{T}_3)\succ 0,
\end{equation}
where $\begin{bmatrix}\tilde{T}_1&\!\!\tilde{T}_2\\\tilde{T}_3&\!\!0\end{bmatrix}\!=$\vspace{-2mm}
\begin{equation*}
    \left.\begin{array}{rl}~&\!\!\left.\!\!\!\!\!\!\!\!\!\!\!\!\!\!\!\!\!\!\begin{array}{ccccc}~~~\scriptstyle{1}&~~~~~~~~\scriptstyle{N_f}&~~~~~~~~~~~~~~~\scriptstyle{N_q}&~~~~~~~~~~~\scriptstyle{N_p}&~~~~\scriptstyle{N_u}\end{array}\right.\\
\!\!\!\!\!\!\left.\begin{array}{c}\scriptstyle{1}\\\scriptstyle{N_f}\\\scriptstyle{N_q}\\\scriptstyle{N_p}\\\scriptstyle{N_p}\end{array}\right.&\!\!\!\!\!\!\!\!\!\!\!\!\!\!
\left[\!\!\!\!\!\begin{array}{cccc|c}
    2\mu & \!\!\!\!\!\!\!(\bar{\mathbf{f}}\!\!-\!\mathbf{D}_f^{\K_0,\hat{\upsilon}}\!\!-\!\!Me\!-\!e\mu)^T &\!\!\!\!\! \!\!\!\!\!\!\!(\mathbf{D}_q^{\K_0,\hat{\upsilon}})^T &\!\!\!\!\!\!\! 0&\!\!\!\!\!0\\
    \ast &\!\!\!\!\!\!\! M\!+\!M^T &\!\!\!\!\!\!\!\!\! -\mathbf{D}_{fp}\tilde{G}^T &\!\!\!\!\!\!\! -\mathbf{D}_{fp}\tilde{S}&\!\!\!\!\!-\mathbf{D}_{fu} \\
    \ast &\!\!\!\!\!\!\! \ast &\!\!\!\!\!\!\!\!\!\!\!\! \tilde{R}\!+\!\H\!(\mathbf{D}_{qp}\tilde{G}^T) &\!\!\!\!\!\!\! \mathbf{D}_{qp}\tilde{S}&\!\!\!\!\!\mathbf{D}_{qu}\\
    \ast &\!\!\!\!\!\!\! \ast &\!\!\!\!\!\!\!\!\!\!\!\! \ast &\!\!\!\!\!\!\! \tilde{S}&\!\!\!\!\!0\\\hline
    0&\!\!\!\!\!\!\!0&\!\!\!\!\!\!\!\!\!\!\!\!\tilde{G}^T&\!\!\!\!\!\!\!\tilde{S}&\!\!\!\!\!0
    \end{array}\!\!\!\!\!\right]\end{array}\right.\!\!\!\!\!\!\!\!\!,
    \end{equation*}
and $( \tilde{S}, \tilde{R}, \tilde{G}) \in \widehat{ \Psi}$, $\mu\in\R$ and $M\in\D^{N_f}$ are additional slack variables. 
It follows that the relaxed RMPC problem can be summarised as:
\begin{equation} \label{finalprob_singleLMI}
\begin{aligned}
\hspace{-2mm}\min\{  {\gamma^2} \!:&([ \hat{K}_0 ~~ \hat{K}],\hat{\upsilon})\!\in\!\mathcal{K}\!\times\mathbb{R}^{N_u},(\ref{cost_LMI}),(\ref{stateupperi0}){\rm~are~satisfied},\\
&(S,R,G),( \tilde{S}, \tilde{R}, \tilde{G})\!\in\! \widehat{\Psi}\}.
\end{aligned}
\end{equation}
The non-linearities appear in \eqref{cost_LMI}, \eqref{stateupperi0} due to terms of the form $\K \mathbf{B}_p \Phi^T$ where $\Phi$ stands for $S$, $G$, $\tilde{S}$ and $\tilde{G}$ . By introducing three new slack variables $Y$, $\tilde{Y}$ and $X$ and using the Elimination lemma derived in~\cite{Georgiou2022}, the problem can be linearised into two LMIs described below in \eqref{linear_cost_LMI} and \eqref{linear_constr_Single_LMI}:
\begin{eqnarray}\label{linear_cost_LMI}
    \begin{bmatrix}
    T_1+\H\left(T_2\bar{K}Y^*\right) & \ast\\
    \left(\mathbf{B}_p T_3-\bar{K}^TT_2^T\right)-XY^* & X+X^T
    \end{bmatrix}\!\!&\succ&\! 0\\\label{linear_constr_Single_LMI}
    \begin{bmatrix}
    \tilde{T}_1+\H\left(\tilde{T}_2\bar{K}\tilde{Y}^*\right) & \ast\\
    \left(\mathbf{B}_p \tilde{T}_3-\bar{K}^T\tilde{T}_2^T\right)-X\tilde{Y}^* & X+X^T
    \end{bmatrix}\!\!\succ\! 0,
\end{eqnarray}
%given all variables defined in \eqref{linear_cost_LMI}, then
%$\mathcal{Z}(\K_0,\K,\hat{\upsilon},\hat{\Delta})\!\le\!{\gamma}^2$ and $\mathcal{F}(\hat{K}_0,\hat{K},\hat{\upsilon},\hat\Delta)\!\le\!\bar{\mathbf{f}}$ for all 
%$\hat{\Delta}\!\in\!\bm\B\pmb{\hat{\Delta}}$  if there exist $([\K_0~\K],\hat{\upsilon})\!\in\!\mathcal{K}\!\times\mathbb{R}^{N_u}$, $(S,R,G), ( \tilde{S}, \tilde{R}, \tilde{G}) \!\in\! \widehat{ \Psi}$, $\mu\!\in\!\R$, $M\!\in\!\D^{N_f}$ and $X \!\in\! \mathbb{R}^{N_n \times N_n}$, with $X$ lower block-triangular with $n\times n$ blocks, to (\ref{linear_cost_LMI}) and the following LMI: %\vspace{-0.0cm}
%\begin{equation}\label{linear_constr_Single_LMI}
%    \begin{bmatrix}
    %\tilde{T}_1+\H\left(\tilde{T}_2\bar{K}\tilde{Y}^*\right) & \ast\\
 %   \left(\mathbf{B}_p \tilde{T}_3-\bar{K}^T\tilde{T}_2^T\right)-X\tilde{Y}^* & X+X^T
%    \end{bmatrix}\!\!\succ\! 0,
%\end{equation}
for any $Y^*\!\in\!\R^{N_n\times(N_z\!+\!1\!+\!N_q\!+\!N_p)},~\tilde{Y}^*\!\in\!\R^{N_n\times(1\!+\!N_f\!+\!N_q\!+\!N_p)}$ and where $\bar{K} := \K X\in\mathcal{K}$ and let $Y^*=\mathbf{B}_pT_3(S^*,G^*)+(T_2\K^*)^T$ and $\tilde{Y}^*=\mathbf{B}_p\tilde{T}_3(\tilde{S}^*,\tilde{G}^*)+(\tilde{T}_2\K^*)^T$. Then \eqref{linear_cost_LMI} and \eqref{linear_constr_Single_LMI} are feasible. The proposed LMI-based RMPC scheme does not restrict the structure of the slack variables $(R,S,G)$ beyond the requirements of $\widehat{\Psi}$, and the reformulation of a single inequality for the constraint signal does not add any additional conservativeness into the problem; see~\cite{Georgiou2022} for more details.

The above formulation shows that the initial non-convex and non-linear RMPC problem can be written as an LMI optimization problem~\cite{Boyd}. $K_0,~K$ and $\pmb{\upsilon}$ can be computed online and applied in the usual receding horizon MPC manner, where the first input of the control sequence $\mathbf{u}$ is applied to the plant, the time window is shifted by 1 sampling instance, the current state is read and the process is repeated.

In this study, a lookup table is built offline to map all the initial state $x_0 \in \mathcal{X}$ to the corresponding $Y^*$ and $\tilde{Y}^*$. Then the offline calculation is implemented to update the initial guesses of $\tilde{S}^*,\tilde{G}^*,\K^*$ once, after which the initial feasible solutions $Y^*(S^*,G^*,\K^*)$ and $\tilde{Y}^*(\tilde{S}^*,\tilde{G}^*,\K^*)$ are obtained and fed online (for more details refer to \cite{Georgiou2022}). 

By following the description that is given for both offline and online controllers, the RMPC strategy that is employed in this paper is summarized in Algorithm 1:
\begin{table}[ht]
\label{ControlStrategy}
\begin{center}
\begin{tabular}{l}
\hline
\textbf{Algorithm 1}: RMPC controller strategy \\
\hline
\\
\textit{Offline calculation:}\vspace{0.05cm}\\
1. Build the lookup table to map all the initial states \\
\quad $x_0 \in \mathcal{X}$ to the corresponding $Y^*$ and $\tilde{Y}^*$.\\
2. Compute the initial feasible solutions $Y^*$ and $\tilde{Y}^*$, by\\
\quad reading the lookup table given the first initial state $x_0$, and fix \\
\quad the value of $Y^*$ and $\tilde{Y}^*$ for the subsequent online calculation\\
\textit{Online calculation:}\vspace{0.05cm} \\
1. Read the current state $x_k$ and set it as initial state $x_0$.\\
\quad Then based on $x_0$, extract the value of $Y^*$ and $\tilde{Y}^*$\\
\quad from the offline calculation.\\
2. Compute the triple ($K_0,K,\upsilon$) through the two LMI\\
\quad procedures outlined in \eqref{linear_cost_LMI} and \eqref{linear_constr_Single_LMI} and apply the first\\
\quad input of the control sequence shown in \eqref{u1}. \\
3. Return to step 1.\\ 
\hline
\end{tabular}
\end{center}
\end{table}

\section{Overall control scheme design}\label{overall:control}
In this section, the issues encountered when adapting the RMPC synthesized in Section~\ref{sec:control} to the nonlinear multi-body model of quarter car SAVGS introduced in Section \ref{sec:2-1} are fully discussed and the overall control scheme is designed.

\subsection{PI control to solve the single link drifting}\label{subsec:PID}
\begin{figure}[ht!]
\centering
\includegraphics[width=0.80\columnwidth]{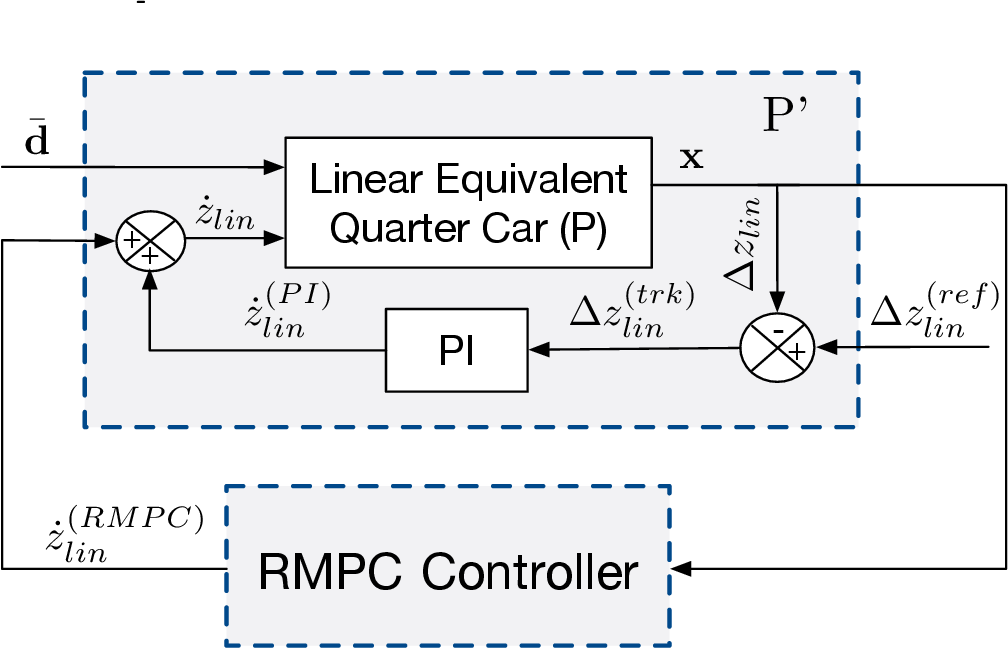}
\caption{RMPC scheme with PI incorporated plant model of linear equivalent quarter car SAVGS, where $\bar{\mathbf{d}}$ corresponds to the stacked vector for disturbance bound, ${P}'$ to the new uncertain system with parallel PI incorporated, and $\Delta z_{lin}^{(ref)}$ to the exogenous reference signal of linear actuator displacement.}
\label{fig1-21}
\end{figure}

At low or zero road disturbance frequencies it is desired for the SL angle ($\Delta z_{lin}$ in the linear equivalent model) to return to the nominal equilibrium state of $\Delta \theta_{SL} = 90\degree$ ($\Delta z_{lin}\,=\,0$ in the linear equivalent model), for example, after traveling over a nonzero road profile and returning to a flat road. However, due to the integrator dynamics between the control input and actuator displacement and the associated drift, this may not occur in the linear equivalent model presented in subsection \ref{subsec:2-B} unless $\Delta z_{lin}$ is explicitly controlled at low or zero frequencies, while a similar argument holds for the nonlinear high fidelity model.
%The linear equivalent model presented in subsection \ref{subsec:2-B} cannot achieve the zero value convergence of linear actuator displacement ($\Delta z_{lin}\,=\,0$) and thereby the single link angle cannot be steered to the nominal equilibrium state of $\Delta \theta_{SL} = 90\degree$ at low or zero frequency, {\color{red}for example, after traveling over a nonzero road profile and returning to a flat road}. 
In previous work~\cite{yu2017quarter} developing SAVGS $H_{\infty}$ control, the transfer function filtering the control output $\dot z_{lin}$ has a free integrator that aims to ensure a zero steady-state tracking error of $\Delta z_{lin}$. In the present work, to avoid overcomplicating the RMPC scheme, a more conventional approach is utilized for the low frequency control, where a proportional-integral (PI) controller is introduced in parallel to the uncertain system developed in subsection \ref{sec:2-3} to ensure zero steady-state error of $\Delta z_{lin}$. As shown in Fig.~\ref{fig1-21}, the exogenous reference signal of the equivalent linear actuator displacement $\Delta z_{lin}^{(ref)} = 0$ is introduced representing the nominal equilibrium of $\Delta\theta_{SL}=90$, and its tracking error ($\Delta z_{lin}^{(trk)}:= \Delta z_{lin}^{(ref)}-\Delta z_{lin}$, where $\Delta z_{lin}=rx$ with $r=[0~0~0~0~1]$ and $x$ the state) is fed back into the PI controller of the form:
\begin{equation}\label{eq.PID}
\dot{z}_{lin}^{(PI)}=K_p \Delta z_{lin}^{(trk)} + K_i\int{\Delta z_{lin}^{(trk)}}dt.
\end{equation}
The parameters $K_p$ and $K_i$ can be tuned based on trial and error. Therefore, the augmented system combines the uncertain system \eqref{1-2} and the PI control \eqref{eq.PID}, which, in continuous-time,
%after discretisation, 
can be described as follows:
\begin{equation}\label{eq:aug1}
\begin{aligned}
\dot{\tilde{x}}(t) &= \tilde{A}\tilde{x}(t) + \tilde{B}_{u}u(t) + \tilde{B}_{p}p(t) + \tilde{B}_{d}d(t),
\end{aligned}
\end{equation}
where $\tilde{x}(t)=[x(t)^T\,\,\,\int\!{(0-\Delta z_{lin})dt}]^T$ is the augmented state and where 
\begin{equation}\label{matrix:aug}
\begin{aligned}
&\tilde{A} = 
\begin{bmatrix}
A+B_uK_pr & B_uK_i \\ 
r & 0\\ 
\end{bmatrix},
&\tilde{B}_u = 
\begin{bmatrix}
B_u\\
0
\end{bmatrix},\\
&\tilde{B}_d = 
\begin{bmatrix}
B_d\\
0
\end{bmatrix},
&\tilde{B}_p = 
\begin{bmatrix}
B_p\\
0
\end{bmatrix},
\end{aligned}
\end{equation}
in which the matrices $A$, $B_{u}$, $B_{d}$, $B_{p}$, are defined in \eqref{matrix_unc}. \\
Then the model after discretisation using a zero-order hold method is defined as:
\begin{equation}\label{eq:aug2}
\begin{aligned}
{x}_{k+1} &= \tilde{A}\tilde{x}_{k} + \tilde{B}_{u}u_{k} + \tilde{B}_{p}p_{k} + \tilde{B}_{d}d_{k},
\end{aligned}
\end{equation}
\begin{rem}Note that the distribution matrices $\tilde{A},\tilde{B}_u,\tilde{B}_p,\tilde{B}_d$, and so on, in \eqref{eq:aug2} are the discretized version of those in \eqref{eq:aug1} and the same notation is used for simplicity.
\end{rem}

\subsection{Design of cost signal}
As explained in Subsection \ref{sec:2-4}, to reduce the vertical body acceleration $\ddot{z}_{s}$ (ride comfort), and vertical tire deflection $\Delta l_{t}$ (road holding), along with minimizing the input command $\dot{z}_{lin}$ (actuator energy consumption), the cost signal in \eqref{eq:dt_system_model} is defined as:
\begin{equation}\label{eq:cost}
\begin{aligned}
z_k=&[w_1\ddot{z}_{s},w_2\Delta l_{t}, w_3\dot{z}_{lin}]^T\\
=&C_zx_k+D_{zu}u_k+D_{zp}p_k+D_{zd}d_k,
\end{aligned}
\end{equation}
with the reference cost in \eqref{eq:costfunction} given by $\bar{z}_{k}=[0\,\, 0\,\, 0]^T$,
where $w_1,w_2$ and $w_3$ are tuneable weights to enhance performance (see Section~\ref{sec:Simulations}). Rewriting $z_k$ in state space form the coefficient matrices are defined as follows:
\begin{equation}
\begin{aligned}
&C_{z~} = 
\begin{bmatrix}
-\frac{c_{eq}}{m_{s}}w_1\,& \frac{c_{eq}}{m_{s}}w_1\,& \frac{k_{eq}}{m_{s}}w_1\,& 0\,& -\frac{k_{eq}}{m_{s}}w_1\,& \\ 
0\,& 0\,& 0\,& w_2\,& 0 \\ 
0\,& 0\,& 0\,& 0\,& 0 
\end{bmatrix},
\\
&D_{zu} = 
\begin{bmatrix}
-\frac{c_{eq}}{m_{s}}w_1\\ 0 \\
w_3
\end{bmatrix}.
\end{aligned}
\end{equation}
For $k=N$ only the terminal state $x_N$ is predicted, thus the terminal cost signal is defined as follows:
\begin{equation}\label{eq:cost_ter}
\begin{aligned}
z_N=&[w_4\ddot{z}_{s},w_5\Delta l_{t}]^T\\
=&\hat{C}_{z}x_N+\hat{D}_{zp}p_N,
\end{aligned}
\end{equation}
where again $w_4$ and $w_5$ are predefined weights. The coefficient matrix $\hat{C}_{z}$ is defined as follows: 
%The matrix Cz_hat is defined by the first two rows of the $C_z$ and multiplied by the appropriate weights.
\begin{equation}
\begin{aligned}
&\hat{C}_{z} = 
\begin{bmatrix}
-\frac{c_{eq}}{m_{s}}w_4\,& \frac{c_{eq}}{m_{s}}w_4\,& \frac{k_{eq}}{m_{s}}w_4\,& 0\,& -\frac{k_{eq}}{m_{s}}w_4\,& \\ 
0\,& 0\,& 0\,& w_5\,& 0 
\end{bmatrix}.
\end{aligned}
\end{equation}
Since the uncertainties and additive disturbances are allowed only in state dynamics, $D_{zp}$, $D_{zd}$ and $\hat{D}_{zp}$ are zero matrices with appropriate dimensions. The dimension of the cost signal does not change but the coefficient matrices change due to augmentation of the PI controller.
The cost signal is thereby redefined for the augmented system as follows:
\begin{equation}\label{eq:cost_overall}
\begin{aligned}&\tilde{z}_k=\tilde{C}_z\tilde{x}_k+\tilde{D}_{zu}u_k,\\
&\tilde{z}_N=\tilde{C}_{zN}\tilde{x}_N
\end{aligned}
\end{equation}
where $\tilde{C}_z = [{C}_z\,\,\,0]$,  
$\tilde{C}_{zN} = [\hat{C}_{z}\,\,\,0]$ and $\tilde{D}_{zu} = {D}_{zu}$.

The robust model predictive control detailed in Section~\ref{sec:control} is synthesized given the derived augmented system in \eqref{eq:aug2} to achieve the (higher frequency) suspension performance objectives, as captured by \eqref{eq:cost_overall}, whilst guaranteeing the (low frequency) tracking property of $\Delta z_{lin}$.

% \subsection{Selection of the disturbance in RMPC design}\label{sec4-B}
% The bound on the disturbance ($\bar{d}$),  which is based on the road profile, is initially chosen as the maximum value of the vertical road velocity. However, when implementing this value in different driving manoeuvres, it is hard to achieve feasibility in the optimization problem. Therefore, the value of the disturbance bound is relaxed in this study and chosen as 0.15\,m/s for all the driving manoeuvres to guarantee a feasible solution. Note that for random roads A, B and C, the disturbances are within the defined range for 100\%, 95.6\% and 85.3\% of the time, respectively. Furthermore, in the case studies presented in Section \ref{sec:Simulations}, it is found that this setting is sufficient to derive well-performing controllers that satisfy all the constraints.

\subsection{Constraint conversion between nonlinear high fidelity model and linear equivalent model}\label{subsec:ConstrainsAssociation}
The nonlinear high fidelity model built in Section \ref{sec:2-4} includes various constraints in the single-link operation, arising from packaging and actuator/gearbox constraints, involving the variables of the single-link angle $\Delta\theta_{SL}$, the single-link velocity $\omega_{SL}$, actuator single link torque $T_{SL}$, PMSM power $P$, and so on; for more details see \cite{Aranaphdthesis}.  %Write something here about $T_{SL}$, PMSM power $P$. 
Since the RMPC is synthesized with the linear equivalent model, the constraints in the nonlinear model should be converted and captured by the linear model also. From the involved variables of the nonlinear model, $\Delta\theta_{SL}$ and $\omega_{SL}$, which are constrained between their minimum and maximum values, $\Delta\theta_{SL}^{(min)}$ and $\Delta\theta_{SL}^{(max)}$, and $\omega_{SL}^{(min)}$ and $\omega_{SL}^{(max)}$, respectively, %and $T_{SL}$ 
are %feasible to 
quantified directly in the linear equivalent model via two conversion functions, respectively. %, corresponding to equivalent actuator displacement $\Delta z_{lin}$ and equivalent actuator velocity $\dot{z}_{lin}$, and the respectively. 
Thus, function $\alpha$ ($=\dot{z}_{lin}/\omega_{SL}$), which has been introduced in Section \ref{sec:model}, can be approximated by a parabolic function with
respect to the single-link angle ($\Delta\theta_{SL}$) \cite{yu2017quarter}, as shown in Fig. 4-top, and is used to convert the nonlinear variable $\omega_{SL}$ to its linear equivalent $\dot{z}_{lin}$. Moreover, function $\beta$ ($=\Delta z_{lin}$), which is the integral of $\alpha$ with respect to the single-link
angle ($\Delta\theta_{SL}$) \cite{yu2017quarter}, as shown in Fig. 4-bottom, is employed to convert the nonlinear variable $\Delta\theta_{SL}$ to its linear equivalent $\Delta z_{lin}$.
\begin{figure}[ht]
\centering
\includegraphics[width=0.9
\columnwidth]{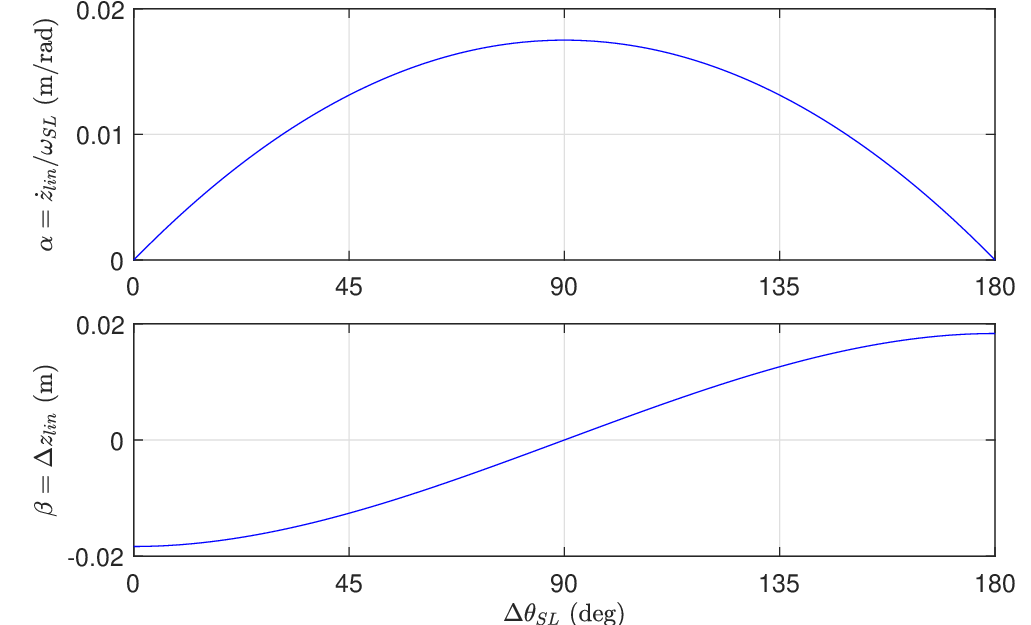}
\caption{Plots of functions $\alpha$ (top) and $\beta$ (bottom) for conversion between the multibody and linear equivalent models.}
\label{fig-conversion}
\end{figure}

The single link torque $T_{SL}$ in the nonlinear model is constrained between ${T}_{SL}^{(min)}$ and ${T}_{SL}^{(max)}$, 
%In the numerical simulation with the nonlinear high fidelity model, constraints of the single link torque $T_{SL}$ are placed between 0\,Nm and 97\,Nm, 
to avoid gearbox backlash effects and to respect the motor continuous torque limit, respectively. Function $\alpha$ is employed again to convert the nonlinear variable $T_{SL}$ to its linear equivalent actuator force $F_{lin}$ ($\alpha=T_{SL}/F_{lin}$ also) \cite{Aranaphdthesis}, where $F_{lin}$ can be calculated by adding the equilibrium (the sprung mass weight) and increment (shown on the right hand side of the first equation in \eqref{motion_eq}) values of the equivalent passive spring force, with the equivalent damper force neglected due to its much smaller magnitude than the equivalent spring force (as verified by simulations). Thus, $F_{lin}$ can be
expressed in terms of the states $\Delta l_s$ and $\Delta z_{lin}$, as follows:
\begin{equation}
F_{lin} = m_sg + k_{eq}(\Delta l_s - \Delta z_{lin}).
\end{equation}
%Rewrite the Cf and Dfu (keq inside), and comment the f_bar in the Numerical simulation!!!
%{\color{blue}The damper force is neglected due to its too small value ($10^3$ less than spring force)} 

Hence, the constraint signal in \eqref{eq:dt_system_model}, capturing all the constraints that are possible to consider in the control synthesis, is defined as:
\begin{equation}\label{constraint}
\begin{aligned}
f_k=&[\Delta z_{lin}, -\Delta z_{lin}, \Delta l_s - \Delta z_{lin}, -\Delta l_s+\Delta z_{lin}, \dot{z}_{lin}, -\dot{z}_{lin}]^T\\
    =&C_fx_k+D_{fu}u_k+D_{fp}p_k+D_{fd}d_k,
\end{aligned}
\end{equation}
where,
\begin{equation}\label{matrix:constraint}
\begin{aligned}
&C_f = 
\begin{bmatrix}
0\,& 0\,& 0\,& 0\,& 1 \\ 
0\,& 0\,& 0\,& 0\,& -1 \\ 
0\,& 0\,& 1\,& 0\,& -1\\
0\,& 0\,& -1\,& 0\,& 1\\
0\,& 0\,& 0\,& 0\,& 0 \\ 
0\,& 0\,& 0\,& 0\,& 0 \\ 
\end{bmatrix},
&D_{fu} = 
\begin{bmatrix}
0 \\ 0 \\ 0 \\ 0 \\ 1 \\ -1  
\end{bmatrix}
\end{aligned}
\end{equation}
The upper bound of constraint signal $f_k$ in \eqref{const_fbar} is defined as:
\begin{equation}\nonumber
\bar{f}_k=\left[
\beta\left(\Delta\theta_{SL}^{(max)}\right),-\beta\left(\Delta\theta_{SL}^{(min)}\right),
\frac{\left(\frac{{T}_{SL}^{(max)}}{{\alpha}}-m_sg\right)}{k_{eq}}, 
\right.
\end{equation}
\begin{equation}\label{eq:fbar}\left.
\frac{\left(m_sg-\frac{{T}_{SL}^{(min)}}{{\alpha}}\right)}{k_{eq}},
\alpha\omega_{SL}^{(max)}, -\alpha\omega_{SL}^{(min)}
\right]^T.
\end{equation}
%where the super subscript (min) and (max) denote the minimum and maximum value of the corresponding constraint signals. 
For $k=N$ only the terminal state $x_N$ is predicted, thus the terminal constraint signal is defined as follows:
\begin{equation}\label{eq:costraint_ter}
\begin{aligned}
f_N=&[\Delta z_{lin}, -\Delta z_{lin}, \Delta l_s - \Delta z_{lin}, -\Delta l_s+\Delta z_{lin}]^T\\
=&\hat{C}_{f}x_N+\hat{D}_{fp}p_N,
\end{aligned}
\end{equation}
The upper bound of terminal constraint signal $\bar{f}_N$ in \eqref{eq:costfunction} is defined as:
\begin{equation}\nonumber
\bar{f}_N=\left[
\beta\left(\Delta\theta_{SL}^{(max)}\right),-\beta\left(\Delta\theta_{SL}^{(min)}\right),
\frac{\left(\frac{{T}_{SL}^{(max)}}{{\alpha}}-m_sg\right)}{k_{eq}}, 
\right.
\end{equation}
\begin{equation}\label{eq:fNbar}\left.
\frac{\left(m_sg-\frac{{T}_{SL}^{(min)}}{{\alpha}}\right)}{k_{eq}}
\right]^T.
\end{equation}

The matrix $\hat{C}_{f}$ is defined by the first four rows of the $C_f$ in \eqref{matrix:constraint}. Similarly to the cost signal in \eqref{eq:cost} and \eqref{eq:cost_ter}, the $D_{fp}$, $D_{fd}$ and $\hat{D}_{fp}$ matrices are equal to zero with appropriate dimension. After augmentation of the PI controller, the dimension of constraint signal stays the same but the coefficient matrices change due to augmentation of the PI controller and thereby defined as follows:
\begin{equation}\label{eq:constraint_overall}
\begin{aligned}
&\tilde{f}_k=\tilde{C}_f\tilde{x}_k+\tilde{D}_{fu}u_k\\
&\tilde{f}_N=\tilde{C}_{fN}\tilde{x}_N
\end{aligned}
\end{equation} where $\tilde{C}_f = [{C}_f\,\,\,0]$,  
$\tilde{C}_{fN} = [\hat{C}_{f}\,\,\,0]$ and $\tilde{D}_{fu} = {D}_{fu}$. The upper bounds $\bar{f}_k$ and $\bar{f}_N$ remain the same.
%write down the other expression including the damper force, but we have proved that the results compared to those without damper force are nearly the same(damper force are very similar))
%In addition to the constraints considered in the control design, power constraints are imposed on the actuator control to limit energy consumption. The constraints imposed in \cite{yu2017quarter}, intentionally limit the single-link actuator power to 500\,W. In this study, the power constraints limit the currents obtained through a simplified computation of electrical power flow to/from the permanent magnetised synchronised motor (PMSM) based on the electromechanical output power and on either motoring mode (500\,W power flow from the PMSM) or generating mode (1500\,W power flow to the PMSM), which can be shown as follows:

\begin{figure}[ht!]
\centering
\includegraphics[width=\columnwidth]{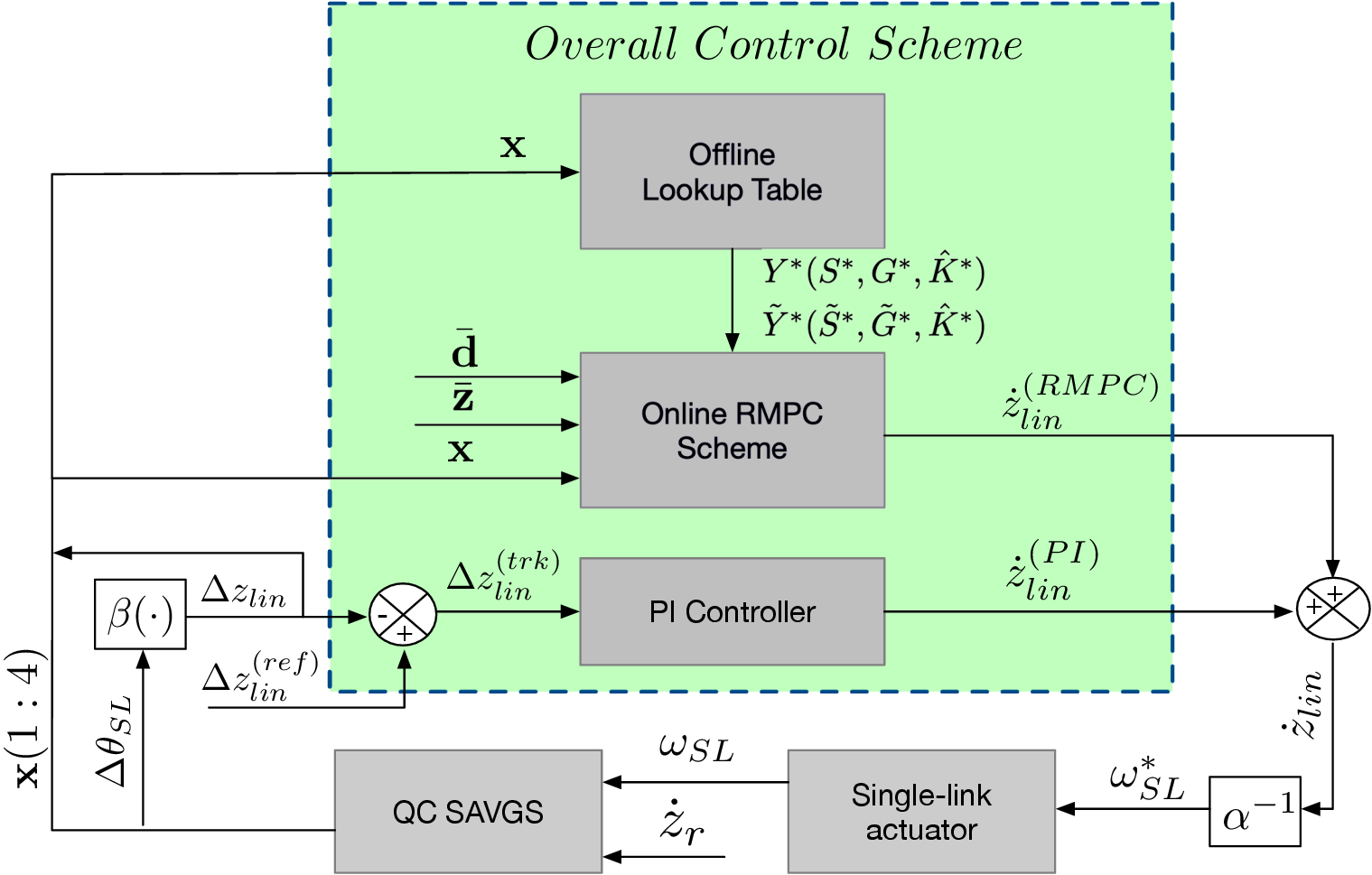}
\caption{Overall simulation block diagram, which contains the high fidelity nonlinear model of quarter car and actuator (`QC SAVGS' and `Single-link actuator', respectively, where $\omega_{SL}^*$ is the reference of $\omega_{{SL}}$ tracked by the actuator) and the proposed coupled closed-loop control scheme. %The RMPC scheme is designed utilizing the linear equivalent uncertain model. 
The `Online RMPC Scheme' uses the linear equivalent uncertain quarter car model and initial feasible sets $(Y^*,\tilde{Y}^*)$ based on offline created lookup tables. %{\color{red}The single link actuator (from $\omega_{{SL}^*}$ to $\omega_{{SL}}$) links the overall control and the mechanical system of the SAVGS quarter car.}
The first four elements of the state $x$ as defined in the subsection \ref{sec:2-2} are measurable at the output of the high-fidelity model and the last state component ($\Delta z_{lin}$), which is also used as input to the PI controller, is computed through the function $\beta$ and the measurable output $\Delta \theta_{SL}$ of the high fidelity model, as explained in subsection \ref{subsec:ConstrainsAssociation}.}
\label{fig1-22}
\end{figure}

It is noted that the minimum and maximum values of the PMSM electrical power $P$, $-\uline{P}$ (power flow to the PMSM) and $\overline{P}$ (power flow from the PMSM), respectively,
\begin{equation}\label{power-constraint}
-\uline{P} \leq P \leq \overline{P},
\end{equation}
are not possible to include as constraints in the RMPC synthesis because of the nonlinearity of such constraints. Instead, they are only imposed in the actuator control model (block `Single-link actuator' in Fig. \ref{fig1-22})
in the simulation of the nonlinear multibody model, via the PMSM 
%d-axis and q-axis currents and voltages ($v_d$,$i_d$), ($v_q$,$i_q$), respectively \cite{Aranaphdthesis}, as follows: \begin{equation}\label{power-constraint}
%-\uline{P} \leq P = \frac{3}{2}(v_di_d + v_qi_q) \leq \overline{P}. 
%\end{equation}
inner (current) and outer (velocity) control loops, by saturating the $dq$ voltage references to the DC-AC converter of the PMSM and the magnetizing $q$ current reference to the inner control loop, respectively, as explained in Sections 4.3 and 4.4 of \cite{Aranaphdthesis}.

The block diagram in Fig.~\ref{fig1-22} shows the proposed overall closed-loop control scheme %of the overall controller (synthesised in Section.\ref{overall:control}) and plant (proposed in Fig.\ref{fig1-21}), 
used for closed-loop simulations with the high fidelity nonlinear model.

%Add the supersubscript (rmpc) after RMPC Controller, and add \dot{z}_{lin} after the summing block before the \alpha^{-1}
\subsection{Benchmark $H_{\infty}$ control scheme}
%The generalised regulator used was the same uncertain linear equivalent model used as in RMPC or the nominal linear equivalent model, there is a PID controller with Hinf or is replaced with the 'M' block as we did before.}

The $H_{\infty}$ control scheme designed in \cite{yu2017quarter} is used as the benchmark scheme in the present work. It uses the same nominal linear equivalent model of \eqref{1-1} and similarly aims to minimize single link displacement tracking error at low frequencies, and vertical body acceleration and tire deflection at higher frequencies. Moreover, the zero convergence of the SL angle is addressed via a free integrator. For more details refer to \cite{yu2017quarter}.

%the interconnection diagram in Fig.\,\ref{fig1-9} is also used to synthesize a conventional $H_{\infty}$ controller, `PALS-$H_{\infty}$', which does not take into account uncertainties, for benchmarking purposes. The same weighting functions as those described in Section \ref{mu-control} for the $\mu$-synthesis control are found to also be beneficial and applied to the $H_{\infty}$ control synthesis.

\section{Numerical Simulations and analysis}\label{sec:Simulations}
In this section, with the nonlinear multi-body quarter car SAVGS model described in 
Section \ref{sec:2-1}, and the control strategies proposed in Sections \ref{sec:control} and \ref{overall:control}, a group of ISO driving maneuvers, containing, i) sinusoidal profile, ii) smoothed bump and hole, and iii) random road class A-C, are tested to evaluate the performance and robustness of the overall synthesized controller, which will be denoted as the RMPC controller. The parameter values of the vehicle (corresponding to a GT car \cite{yu2017quarter}), SAVGS actuator \cite{Aranaphdthesis,yu2017quarter}, structured uncertainty, system constraints, disturbance bounds, reference cost, weights of the cost function, prediction horizon , and PI control gains are shown in Table~\ref{tab1-2}.
\begin{table}[htb!]
\centering
\caption{%Main Parameters of SAVGS quarter car, and linear equivalent model and weight of cost in RMPC synthesis
Main Parameters of SAVGS quarter car nonlinear multi-body and linear equivalent models, and RMPC and PI control, used in numerical simulations}\label{tab1-2}
\scalebox{1}{\begin{tabular*}{1\columnwidth}{@{\extracolsep{\fill}}l @{\extracolsep{\fill}} c@{\extracolsep{\fill}}
c}
\hline
\hline
%Parameter 
Description & Symbol & Value \\
 \hline
weight of sprung mass & $m_s$ & 320\,kg\\
weight of unsprung mass & $m_u$ & 49\,kg\\
tire's radial stiffness & $k_t$ & 275000\,N/m\\
tire's radial damping & $c_t$ & 300\,kg\\
equivalent spring stiffness & $k_{eq}$ & 59987\,N/m\\
actuator maximum power limit & $\bar{P}$ & 500\,W\\
actuator minimum power limit & $-\uline{P}$ & -1500\,W\\
nominal suspension damping & $c_{eq}^{(nom)}$ & 2087.4\,Ns/m \\
suspension damping deviation & $c_{eq}^{(dev)}$ & 208.74\,Ns/m\\
upper bound of SL angle & $\Delta\theta_{SL}^{(max)}$ & %2.792\,rad
160\,deg\\
lower bound of SL angle & $\Delta\theta_{SL}^{(min)}$ & %0.39\,rad
20\,deg\\
upper bound of continuous SL torque & $T_{SL}^{(max)}$ & 97\,Nm\\
lower bound of continuous SL torque & $T_{SL}^{(min)}$ & 0\,Nm\\
upper bound of SL angular velocity & $\omega_{SL}^{(max)}$ & 13\,rad/s\\
lower bound of SL angular velocity & $\omega_{SL}^{(min)}$ & -13\,rad/s\\
upper bound of road disturbance & $\bar{d}_{k}$ & 0.15\,m/s\\
%reference cost signal & $\bar{z}_{k}$ & $[0\,\, 0\,\, 0]^T$\\
vertical body acceleration
weight & $w_1$ & $\sqrt{10}$\\
vertical tire deflection weight & $w_2$ & $\sqrt{10}$\\
actuator energy consumption weight & $w_3$ & $10\sqrt{6}$\\
vertical body acceleration terminal weight & $w_4$ & 20\\ vertical tire deflection terminal weight & $w_5$ & 20\\
prediction horizon & $N$ & 5\\
PI proportional gain & $K_p$ & 5\\
PI integral gain & $K_i$ & 1\\
\hline
\hline
\end{tabular*}}
\end{table}
\vspace{2ex}

In addition to what was done in \cite{yu2017quarter}, which imposed the same single-link actuator power limits of 500\,W for both directions of actuator power flow, in this work $\overline{P} \leq \uline{P}$ to allow for higher power flows when generating as compared to motoring.

%\subsection{Selection of the disturbance in RMPC design}\label{sec4-B}
The bound on the disturbance ($\bar{d}_k$),  which is based on the road profile, is initially chosen as the maximum value of the vertical road velocity. However, when implementing this value in different driving manoeuvres, it is hard to achieve feasibility in the optimization problem. Therefore, the value of the disturbance bound is relaxed in this work and chosen as 0.15\,m/s for all the driving manoeuvres to guarantee a feasible solution. Note that for random roads A, B and C, the vertical road velocities are within the defined range for 100\%, 95.6\% and 85.3\% of the time, respectively. Furthermore, in the case studies presented in this section, %\ref{sec:Simulations}, 
it is found that this setting is sufficient to derive well-performing controllers that satisfy all the constraints.

Initial feasible triples $Y^*(S^*,G^*,\hat{K}^*)$ and $\tilde{Y}^*(\tilde{S}^*,\tilde{G}^*,\hat{K}^*)$ are computed offline utilizing the proposed method in ~\cite{Georgiou2022}, where the initial state $x_0$ is assumed to be within a bounded set $\mathcal{X}$ which includes all the possible values that the state can take, based on the geometry and dynamics of the system.
\begin{rem}
Alternatively, the initial parameters $ S^*,G^*,\tilde{S}^*,\tilde{G}^*$ and $\hat{K}^*$ can be initially set to zero at $t=0$ and then at every sampling time their values can be updated by the slack variables $(S,G,\tilde{S},\tilde{G})$ and control value $K^*$ computed by the online RMPC problem presented in~\eqref{finalprob_singleLMI}. Note that this strategy does not guarantee initial feasibility of the RMPC problem, however, if the problem is solvable at $t=0$, the upcoming optimal control problems presented in ~\eqref{finalprob_singleLMI} have a solution.
\end{rem}
%...the initial state is obtained from the RMPC offline calculation~\cite{Georgiou2022}.

The weights of the RMPC cost signal ($w_1$, $w_2$ and $w_3$) are tuned using trial and error, while the terminal cost weights ($w_4$ and $w_5$) are chosen larger than $w_1$ and $w_2$, which enhance convergent performance and stability. %to the invariant set.
To maintain the computational time of the online control problem within the discretized sampling time, the prediction horizon is set as $N=5$. Also, the PI parameters are tuned by trial and error to achieve good tracking of the SL angle reference at low frequencies. Note that a) the tuning of the PI controller is done first, without the RMPC being in place, and b) the tuning of the RMPC weights is done with the PI controller in place and with its parameters already tuned. This is to achieve the higher frequency suspension performance requirements of ride comfort and road holding simultaneously with the low-frequency SL angle tracking requirements, but without the RMPC interfering much with the low-frequency tracking of the SL angle and vice versa.

\subsection{Simulation With Harmonic Road}
\begin{figure}[ht]
\begin{center}
\includegraphics[width=9.1cm]{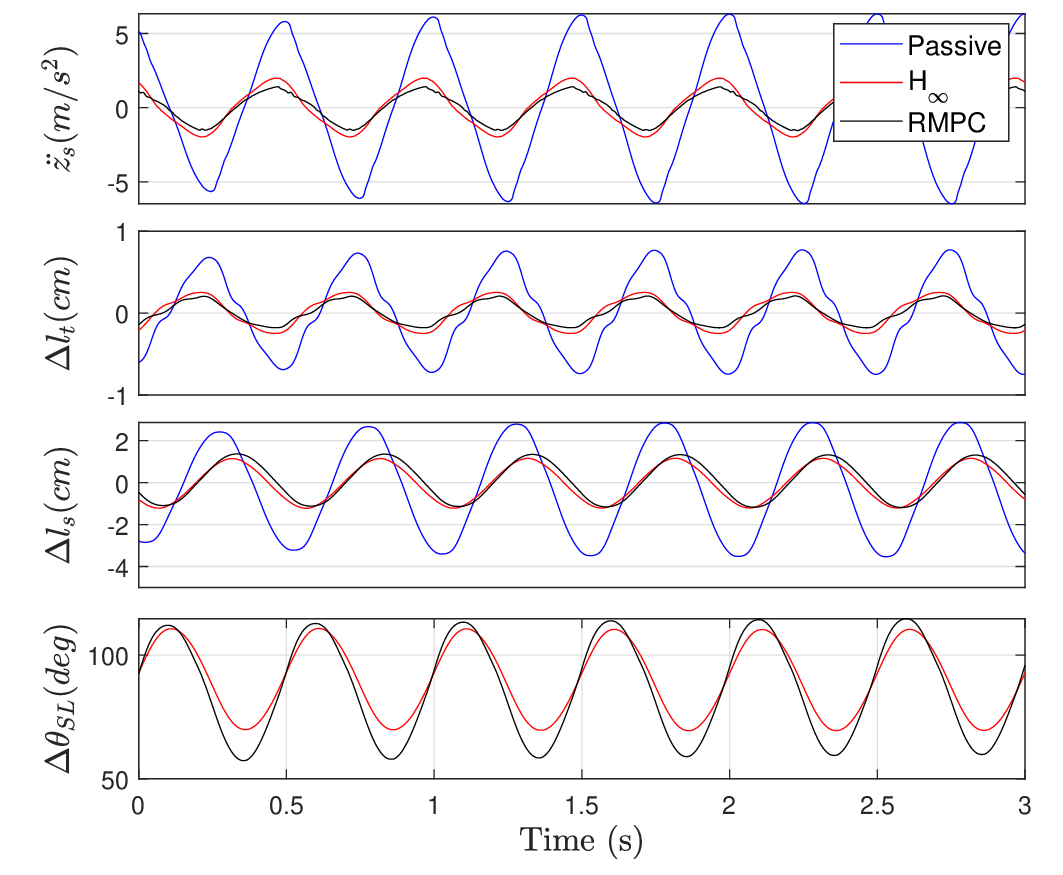}
\vspace{-2mm}
\caption{Numerical simulation results with the nonlinear high fidelity model (top to bottom): the sprung mass acceleration ($\ddot{z}_s$), tire deflection increment ($\Delta l_t$) and suspension deflection increment ($\Delta l_s$) with respect to the initial nominal equilibrium state (SL at the $\Delta\theta_{SL}=90\degree$ position), and single-link angle ($\Delta\theta_{SL}$), when the quarter-car SAVGS is undergoing a harmonic road profile with 2\,Hz frequency and 2.75\,cm peak-to-peak amplitude, for the passive suspension, and $H_{\infty}$ and RMPC active suspension control. %Note that the initial state is defined as the car driving at 10\,km/h constant speed and with SL angle fixed at 90$\degree$.
}

\label{fig1-3}
\vspace{-2mm}
\end{center}
\end{figure}

\begin{figure}[ht]
\begin{center}
\includegraphics[width=9.1cm]{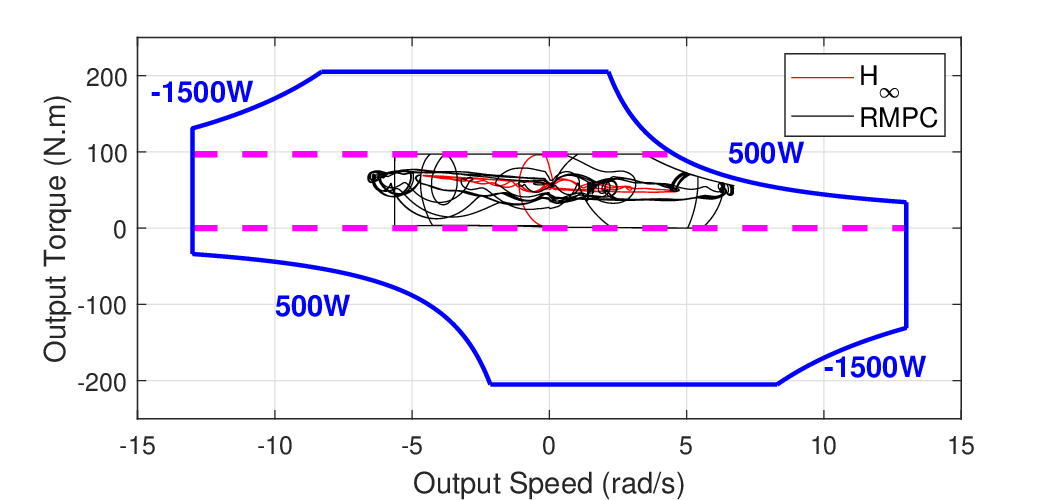}
\vspace{-2mm}
\caption{SAVGS actuator output torque ($T_{SL}$) vs. output speed ($\omega_{SL}$) characteristics from simulation with the nonlinear high fidelity model, when the quarter-car SAVGS is undergoing a sinusoidal road profile with 2\,Hz frequency and 2.75\,cm peak-to-peak amplitude, for different cases of active suspension control, with actuator peak and continuous limit boundaries shown in blue solid and magenta dashed lines, respectively.}
\label{fig1-4}
\end{center}
\end{figure}

Fig. \ref{fig1-3} presents the time response results for the body acceleration, tire deflection, suspension travel and single-link angle of the quarter-car in response to a sinusoidal road disturbance of frequency 2 Hz, for the passive, $H_{\infty}$ and RMPC controllers. It can be seen that the $H_{\infty}$ controller can significantly attenuate the performance objectives, with a 67\% and 64\%\, root mean square value (RMS) drop in the sprung mass acceleration and the tire deflection, respectively, as compared to the passive suspension. The RMPC controller achieves further improvements by reducing further body acceleration by 18.7\% and tire deflection by 19.8\%, as compared to the $H_{\infty}$ controller.
%The RMPC controller has a similar performance to that of the $H_{\infty}$ controller with marginally larger mitigation in terms of body acceleration (74\%) and tire deflection(70\%), as compared to the passive suspension. 
Moreover, Fig.~\ref{fig1-4} indicates that the RMPC control scheme utilizes a larger part of the available actuator output torque-speed operating range than the $H_{\infty}$ scheme does. It does so while still satisfying all the hard (torque and speed) and soft (power) constraints of the SAVGS actuator, which also reflects on the slightly larger suspension travel and single-link angle with the RMPC than with the $H_{\infty}$ scheme in Fig. \ref{fig1-3}. 

%Change the third vertical labels in both figs!!!

\subsection{Simulation With Smoothed Bump and Hole}
Speed bumps or humps are common in some roadways and are normally approximated as a raised cosine shape. The mathematical representation of the wheel road height while running over a standard laterally uniform bump with 0.0275\,m height and 1.4\,m length, and after a distance of 4.15\,m a hole with 0.0275\,m height and 1.4\,m length, is expressed as follows:
\begin{equation}\label{eq:bumphole}
    h_{bh}(x)=\left\{\begin{array}{ll} 
    h_c(1-\cos(\frac{2\pi x}{1.4})),&~0<x\le1.4 \\
    h_c(-1+\cos(\frac{2\pi (x-5.55)}{1.4})),&~5.55<x\le6.95\\ 
    0, &~\text{elsewhere} 
    \end{array}\right.
\end{equation}
where $h_c=0.01375$.

\begin{figure}[ht]
\begin{center}
\includegraphics[width=9.1cm]{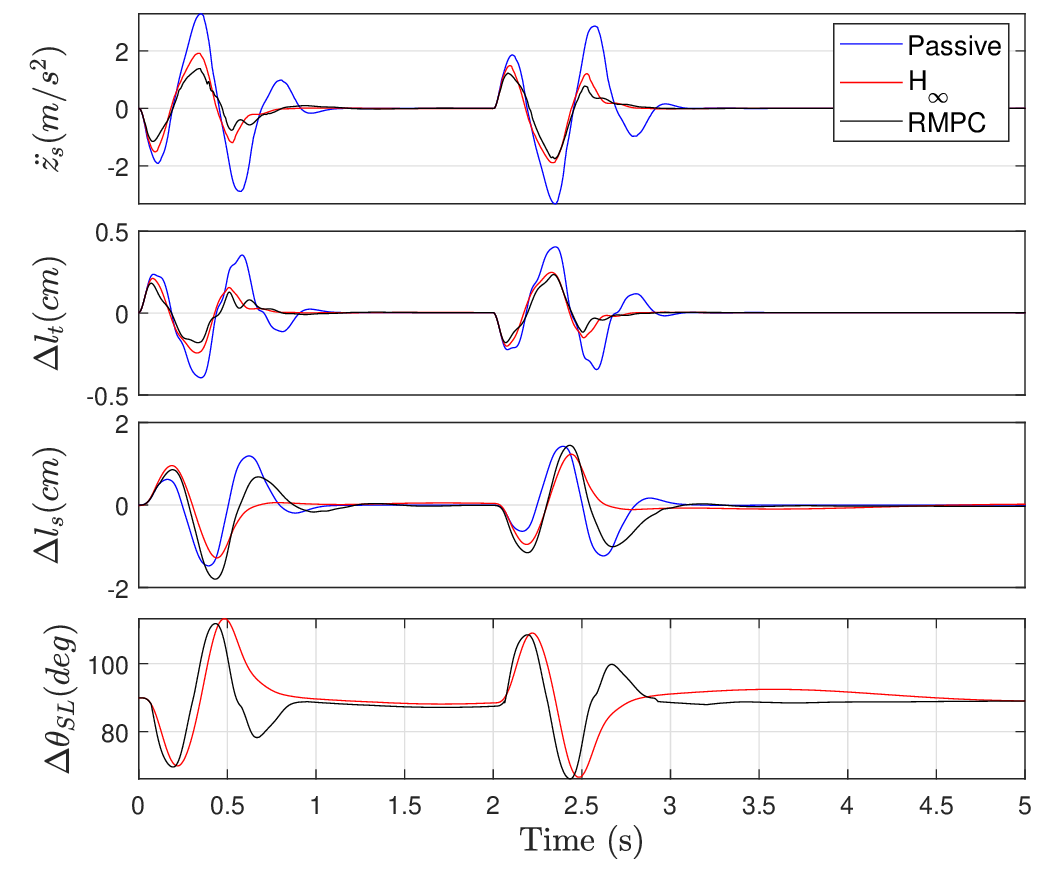}
\vspace{-2mm}
\caption{Numerical simulation results with the nonlinear high fidelity model (top to bottom): the sprung mass acceleration ($\ddot{z}_s$), tire deflection increment ($\Delta l_t$) and suspension deflection increment ($\Delta l_s$) with respect to the initial nominal equilibrium state (SL at the $\Delta\theta_{SL}=90\degree$ position), and single-link angle ($\Delta\theta_{SL}$), when the quarter-car SAVGS is running at 10\,km/h driving speed over a smoothed bump (0-2\,s) and hole (2-4\,s), both with 2.75\,cm in road height and 1.4\,m length (profile in \eqref{eq:bumphole}), for the passive suspension, and $H_{\infty}$ and RMPC active suspension control. %Note that the initial state is defined as the car driving at 10\,km/h constant speed and with SL angle fixed at 90$\degree$.
}
\label{fig1-5}
\vspace{-5mm}
\end{center}
\end{figure}

\begin{figure}[ht]
\begin{center}
\includegraphics[width=9.1cm]{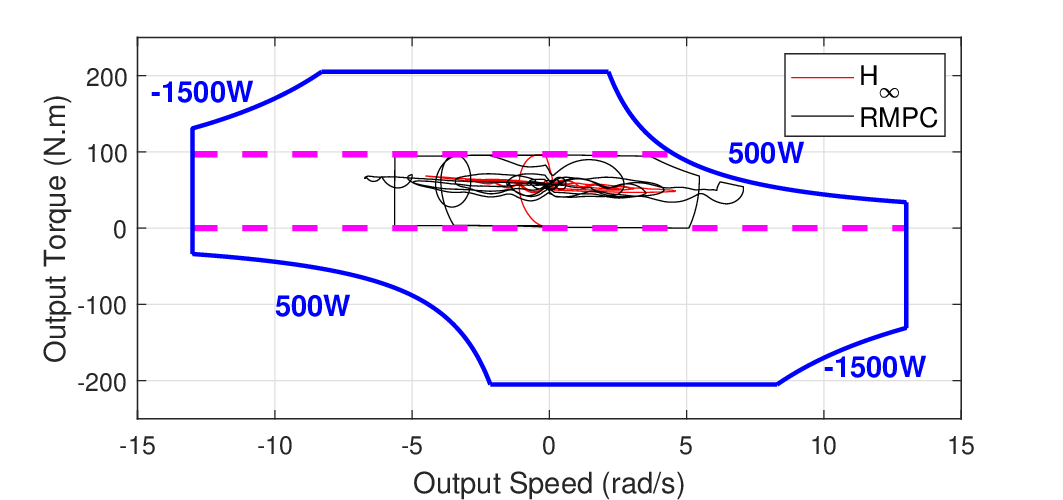}
\vspace{-2mm}
\caption{SAVGS actuator output torque ($T_{SL}$) vs. output speed ($\omega_{SL}$) characteristics from simulation with the nonlinear high fidelity model, when the quarter-car SAVGS is running  at 10\,km/h driving speed over a smoothed bump and hole road profile as in \eqref{eq:bumphole}, for different cases of active suspension control, with actuator peak and continuous limit boundaries shown in blue solid and magenta dashed lines, respectively.}
\label{fig1-6}
\vspace{-2mm}
\end{center}
\end{figure}

Numerical simulation results at a forward speed of 10\,km/h over the road profile in \eqref{eq:bumphole} with the passive and SAVGS-retrofitted quarter car with different active control cases, are shown in Figs. \ref{fig1-5} and \ref{fig1-6}. %The distance between bump and hole is 4.15\,m. 
Similarly to the sinusoidal road cases, the $H_{\infty}$ scheme outperforms the passive suspension. In turn the RMPC outperforms the $H_{\infty}$ active suspension with a 27.68\% and 25.95\% further reduction in terms of sprung mass acceleration and tire deflection peak values, respectively. It can also be observed that the PI controller of the overall RMPC scheme successfully restores the single-link angle to its equilibrium value (SL at the $\Delta\theta_{SL}=90\degree$ position) after each of the bump and hole events. Furthermore, as it can be seen in Fig \ref{fig1-6}, the actuator constraints are satisfied by both control schemes, but the actuator output torque-speed operating range is more widely used in the RMPC scheme.

\subsection{Simulation with Random Road}
The ISO random roads are used to simulate road unevenness. The measured vertical surface data of different road profiles, such as streets, highways as well as off-road terrain are usually described in terms of their power spectral density (PSD), which is defined in the frequency domain as follows~\cite{ISO_8608-2016}:
\begin{equation}
\begin{aligned}
G_{d}(n)=10^{-6}\cdot2^{2k} (\frac{n}{n_0})^{-\bar{\omega}},
\end{aligned}
\end{equation}
where %variable 
$n\in [n_{min},\,n_{max}]$ is the spatial frequency (cycles per meter), $k=2$ to $9$ corresponds to the road roughness classes $A$ to $H$, respectively, $n_0=0.1$\,cycles/m is the reference spatial frequency, and $\bar{\omega}=2$ is a constant. %The spatial frequency range [$n_{min}$,\,$n_{max}$] is suggested as [0.01,\,10] for the general on-road vehicles.

The random road profile in the time domain is obtained as the addition of a series of harmonic signals with varying amplitudes and spatial frequencies as follows \cite{Aranaphdthesis}:
\begin{equation}
\begin{aligned}
&h_{ran}(x)=\sum_{n_i = n_{min}}^{n_{max}}A_i cos(2\pi n_i x + \phi_i), \\
%&G_d(n)\mid_{n=n_i} = \lim_{\Delta n \to 0}\frac{A_{i}^{2}/2}{\Delta n},
&A_i=\sqrt{2\,\Delta n\,G_d(n_i)}=2^{(k+1/2)}10^{-3}\sqrt{\Delta n}\left(\frac{n_0}{n_i}\right),
\end{aligned}
\end{equation}
in which $h_{ran}(x)$ is the random road height. Sinusoidal components are related to the spatial frequency $n_i$ and the random phase $\phi_i$, which is distributed uniformly over the range (0, 2$\pi$). $\Delta n$ is the spatial frequency step, and $A_{i}$ is the amplitude corresponding to the spatial frequency $n_i$ (for more details refer to \cite{Aranaphdthesis}).

The parameters of the random road profiles for simulations with the nonlinear high fidelity model are selected as follows: the spatial frequency range [$n_{min}$,$n_{max}$]\,=\,[0.01,10] cycles/m for the general on-road vehicles, the road length $L$=1\,km, $\Delta n$ = 0.001\,cycles/m since $\Delta n\leq1/L$ should be satisfied, and $k$ takes one of three values to correspond to three different cases of highway road unevenness, a good quality highway (class A, $k=2$), an average quality road (class B, $k=3$), and a poor quality road (class C, $k=4$). The simulations are conducted with a forward speed of $100$\,km/h of the SAVGS-retrofitted quarter-car over each of the three road sections, and are used to validate the improvement of the ride comfort and road holding. 

%In the numerical simulation environment, the SAVGS-retrofitted quarter car is driven on a highway road with a forward speed of $100$\,km/h. Different road profiles (of the same unevenness) against travelled distance are generated for the wheel. In the present work, three 10 km long road sections have been generated and used to validate the improvement of the ride comfort and road holding in typical uneven road surface conditions corresponding to a good highway (class A), an average quality road (class B), and a poor quality road (road C).

\begin{figure}[ht]
\centering
\includegraphics[width=0.95\columnwidth]{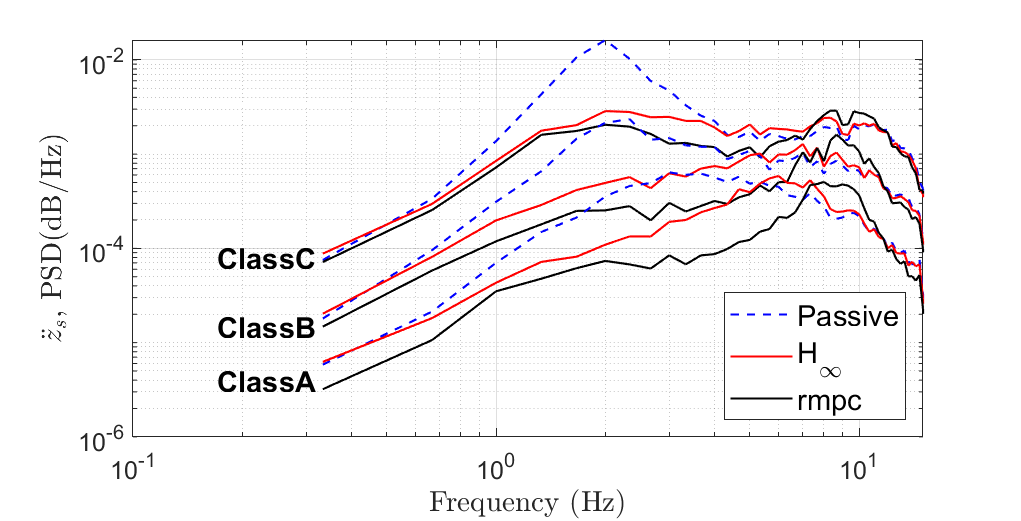}
\includegraphics[width=0.95\columnwidth]{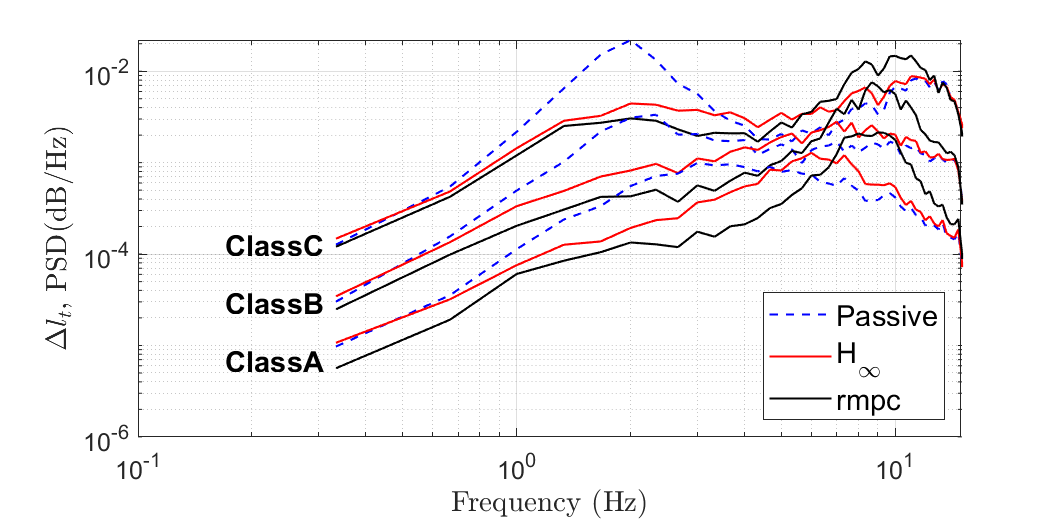}
\caption{Numerical simulation results: the PSD estimate of the sprung mass acceleration (top) and tire deflection (bottom), when the quarter car SAVGS is travelling with a forward speed of $100$\,km/h over random roads (Classes A-C) for the passive suspension and different cases of active suspension control.}
\label{fig1-7}
\end{figure}

The nonlinear simulation power spectral densities (PSDs) of the sprung mass acceleration and the tire deflection over the random roads (classes A-C) are shown in Fig. \ref{fig1-7}. The $H_{\infty}$ and RMPC controllers both give a notably improved performance in terms of ride comfort and road holding at around human-sensitive frequencies (1-6\,Hz) as compared to the passive case. It can also be observed that the RMPC outperforms $H_{\infty}$ in terms of sprung mass acceleration attenuation at almost all frequencies, and in terms of tire deflection reduction at frequencies below approximately 7\,Hz. Despite this tire deflection deterioration above approximately 7\,Hz with RMPC, the sprung mass acceleration and tire deflection RMS values detailed in Table \ref{tab1-1} demonstrate that overall there is improvement in both metrics by RMPC as compared to both the passive and $H_{\infty}$ active suspension. Thus, the RMPC reduces the sprung mass RMS acceleration by 5.6\% (road C) up to 12.3\% (road A) and the tire RMS deflection by 0.4\% (road C) up to 7\% (road A), as compared to $H_{\infty}$, with the reductions as compared to the passive suspension being even larger.

\begin{figure}[ht]
\begin{center}
\includegraphics[width=8.4cm]{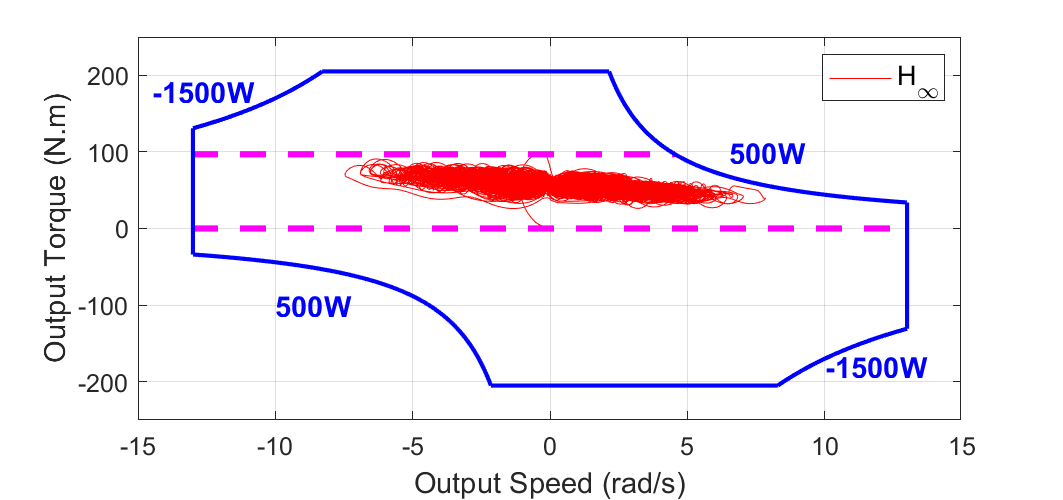}
\includegraphics[width=8.4cm]{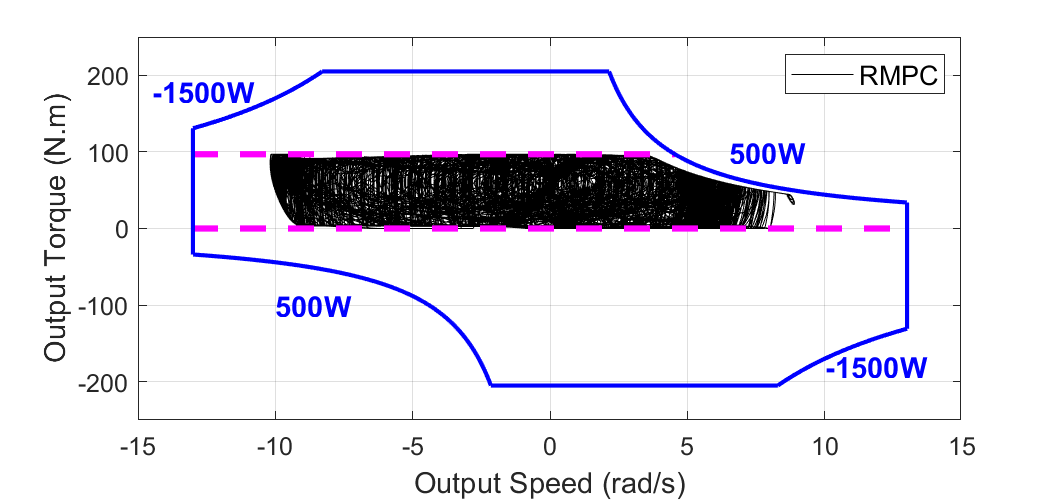}
\vspace{-2mm}
\caption{SAVGS actuator output torque ($T_{SL}$) vs. output speed ($\omega_{SL}$) characteristics from simulation with the nonlinear high fidelity model, when the quarter-car SAVGS is traveling with a forward speed of $100$\,km/h over a random road profile (class C) for $H_{\infty}$ control (top) and RMPC (bottom), with actuator peak and continuous limit boundaries shown in blue solid and magenta dashed lines, respectively.}
\label{fig1-8}
\end{center}
\end{figure}

The output torque-speed operating points for the actuators are plotted in Fig.~\ref{fig1-8} alongside the power, torque and speed constraint envelope. Out of the three different random road cases, the most power-consuming event, for the poor quality random road C, is presented. Results show that the proposed RMPC control utilizes the actuator capabilities fully: the operating torque-speed points approach the boundaries without exceeding them. The $H_{\infty}$ control only employs part of the actuator capability leading to the conservative performance in terms of ride comfort and road holding, as discussed in Fig. \ref{fig1-7}.

\begin{table}[ht]
\setlength\tabcolsep{3pt}
\centering
\caption{RMS values of $\ddot{z}_{s}$\, and $\Delta l_{t}$ with passive suspension and two different controllers for random road Classes A-C. Columns denote passive suspension, $H_{\infty}$, RMPC, and percentage improvement of RMPC with respect to $H_{\infty}$ (last column). The numbers in the brackets denote the percentage of improvement of the two active control methods with respect to the passive suspension.}\label{tab1-1}
\vspace{-2mm}
\begin{tabular*}{1\columnwidth}{ @{\extracolsep{\fill}}c|ccccc}\hline\hline
& Symbol & Passive & $H_{\infty}$ & $RMPC$\\\hline
& $\ddot{z}_{s}$ & 1.910 & 1.600 (16\%) & 1.510 (21\%) & 5.6\%\\ %5.8\%\\
C& $\Delta l_{t}$ & 0.281 & 0.270 (4.1\%) & 0.269 (4.5\%) & 0.4\%\\ \hline
&$\ddot{z}_{s}$ & 1.060 & 0.950 (9.3\%) & 0.870 (16.8\%) & 8.4\%\\ %8.3\%\\
B&$\Delta l_{t}$ & 0.144 & 0.147 (-1.9\%)& 0.142 (2.1\%)& 3.4\%\\ \hline
&$\ddot{z}_{s}$ & 0.630 & 0.570 (8.6\%)& 0.500 (19.5\%) &12.3\%\\ %11.8\%\\
A& $\Delta l_{t}$ & 0.083 & 0.086 (-2.2\%) & 0.080 (2.5\%) & 7.0\%\\
\hline\hline
\end{tabular*}
\end{table}

\section{CONCLUSIONS}\label{sec:conclusion}
The recently proposed mechatronic suspension of the Series Active Variable Geometry Suspension (SAVGS) \cite{arana2016series,arana2017series,yu2017model} is investigated in the application to a quarter car, with suspension damping nonlinearities and road profiles taken into consideration in the suspension control design as uncertainties and bounded disturbances, respectively, revealing promising potential in terms of ride comfort and road holding improvement.

An LMI-based robust model predictive control (RMPC) scheme with an uncertain system description is proposed as the control design that can effectively improve the road holding and ride comfort performance at the human sensitive frequency range (1-6\,Hz), as compared to the passive suspension and a benchmark $H_{\infty}$ control, while system stability and constraint satisfaction are preserved. %by the offline Algorithm~1 proposed in~\cite{Georgiou2022}, see Fig.~\ref{fig1-8}. 
%As compared to a benchmark $H_{\infty}$ control, the RMPC control provides a substantial performance improvement and ensures hard constraint satisfaction see Fig. \ref{fig1-7} and Fig. \ref{fig1-8} respectively.
In particular, the proposed RMPC scheme provides in simulations of ISO maneuvers of sinusoidal road, smoothed bump and hole, and random road A, B and C with a high fidelity model of the quarter-car with SAVGS decent performance improvement as compared to $H_{\infty}$, with 18.7\%, 19.8\% and 12.3\% in terms of rms body acceleration (ride comfort), respectively. Improvements over the $H_{\infty}$ control in terms of tire deflection (road holding) are also provided. Overall, the results illustrate the effectiveness of the proposed control method to extract further performance out of the recently proposed SAVGS architecture, as compared to previous control methods, utilizing the SAVGS actuator at its limits of capability.
\bibliographystyle{IEEEtran}
\bibliography{reference}

%%%%%%%%%%%%%%%%%%%%%%%%%%%%%%%%%%%%%%%%%%%%%%%%%%%%%%%%%%%%%%%%%%%%%%%%%%%%%%%%

\end{document}